\documentclass[preprint,superscriptaddress,nofootinbib]{revtex4}
\usepackage[dvips]{graphicx}
\usepackage[usenames]{color}
\usepackage{setspace}
\usepackage{amsmath}
\usepackage{amssymb}
\usepackage{amsthm}
\usepackage{epsf}
\usepackage{epsfig}

\begin{document}

\newcommand{\GeV}{{\text{GeV}}}
\newcommand{\Hz}{{\text{Hz}}}
\newcommand{\eff}{{\text{eff}}}
\newcommand{\SM}{{\text{SM}}}
\newcommand{\BSM}{{\text{BSM}}}

\renewcommand{\topfraction}{0.8}


\preprint{UT-HET 112}

\title{
  Synergy between measurements of the gravitational wave
  and the triple Higgs coupling in probing first order 
  phase transition
}
\author{Katsuya~Hashino} 
\email{hashino@jodo.sci.u-toyama.ac.jp}
\affiliation{
  Department of Physics,
  University of Toyama, 3190 Gofuku, Toyama 930-8555, Japan
}
\author{Mitsuru~Kakizaki}
\email{kakizaki@sci.u-toyama.ac.jp}
\affiliation{
  Department of Physics,
  University of Toyama, 3190 Gofuku, Toyama 930-8555, Japan
}
\author{Shinya~Kanemura}
\email{kanemu@sci.u-toyama.ac.jp}
\affiliation{
  Department of Physics,
  University of Toyama, 3190 Gofuku, Toyama 930-8555, Japan
}
\author{Toshinori~Matsui} 
\email{matsui@jodo.sci.u-toyama.ac.jp}
\affiliation{
  Department of Physics,
  University of Toyama, 3190 Gofuku, Toyama 930-8555, Japan
}

\begin{abstract}
  
Probing the Higgs potential and new physics behind the electroweak symmetry
breaking is one of the most important issues of particle physics.
In particular, nature of electroweak phase transition is essential for understanding 
physics at the early Universe, such that the strongly first order phase transition is 
required for a successful scenario of electroweak baryogenesis. 
The strongly first order phase transition is expected to be tested by precisely measuring 
the triple Higgs boson coupling at future colliders like the International Linear Collider. 
It can also be explored via the spectrum of stochastic gravitational waves to be 
measured at future space-based interferometers such as eLISA and DECIGO. 
We discuss complementarity of both the methods in testing the strongly first order phase transition 
of the electroweak symmetry in models with additional isospin singlet scalar fields 
with and without classical scale invariance. 
We find that they are synergetic in identifying specific models of 
electroweak symmetry breaking in more details. 

\end{abstract}


\maketitle



\cleardoublepage

\section{Introduction}
\label{sec:introduction}

The discovery of the Higgs boson ($h$), whose mass is 125 GeV,
at the Large Hadron Collider (LHC) Run-I~\cite{Aad:2012tfa,Chatrchyan:2012xdj}
and subsequent measurements of properties of the Higgs 
boson~\cite{Khachatryan:2014jba,Aad:2015gba} have 
established the standard model (SM) of particle physics 
as a low energy effective theory 
at the electroweak scale. 
Nevertheless, the details of the Higgs sector 
including the shape of the Higgs potential
are still unknown.
The Higgs sector of the SM with one isospin doublet scalar field is constructed based merely on minimality, 
and not guided by any fundamental principle. 
Many models with extended Higgs sectors can also satisfy
current experimental data, and need to be distinguished in the future. 
In addition to the ambiguity of the Higgs potential, several 
phenomena that require physics beyond the SM (BSM) have been reported.
These include baryon asymmetry of the Universe (BAU)~\cite{Agashe:2014kda}, 
the existence
of dark matter, cosmic inflation and neutrino oscillations.
Therefore, we are obliged to construct models of BSM 
to tackle such problems.
It should be noticed that
many BSM models predict different classes of the Higgs sector
from the viewpoint of the number of Higgs multiplets and
imposed symmetries.
Moreover, the essence of the Higgs boson 
is directly connected to a BSM paradigm such as 
supersymmetry (SUSY), dynamical symmetry breaking, classical scale invariance (CSI).
Therefore, hints of new physics can be obtained by exploring
the properties of the Higgs sector.

One of the prominent examples of new physics accessible 
through probing the Higgs potential is 
electroweak baryogenesis (EWBG) ~\cite{Kuzmin:1985mm}.
Successful scenarios of EWBG require strongly first order phase
transition (1stOPT) for the electroweak symmetry breaking (EWSB).
With the mass of the discovered Higgs boson being $125~\GeV$,
strongly 1stOPT cannot occur in the framework of the SM Higgs sector.
However, by extending the Higgs sector with additional scalar fields,
strongly 1stOPT can be easily 
realized~\cite{Funakubo:1993jg,Cline:1996mga,Kanemura:2004ch,Funakubo:2005pu,Profumo:2007wc,Noble:2007kk,Aoki:2008av,Kanemura:2011fy,Gil:2012ya,Tamarit:2014dua,Kanemura:2014cka,Fuyuto:2014yia,Profumo:2014opa,Blinov:2015vma,Fuyuto:2015vna,Karam:2015jta,Kakizaki:2015wua}.
In order to unveil the mechanism of EWBG,
one needs to identify the shape of the Higgs potential 
and the dynamics of the EWSB.

It is known that the condition for the 1stOPT for EWSB leads to
the deviation in the triple Higgs boson coupling (the $hhh$ coupling), 
which is typically larger than $10\%$ 
in renormalizable extended Higgs sectors~\cite{Kanemura:2004ch,Noble:2007kk,Aoki:2008av,Kanemura:2011fy,Tamarit:2014dua,Kanemura:2014cka,Kakizaki:2015wua}.
In addition, the deviation in the $hhh$ coupling and the 1stOPT due to
dimension-six operator in the Higgs sector have been discussed in Refs.~\cite{Grojean:2004xa,Grinstein:2008qi}. 
One method to investigate such strong 1stOPT is to measure the $hhh$ coupling
at future high-energy electron-positron colliders.
It seems to be challenging to measure the $hhh$ coupling with such an accuracy 
even at the high luminosity LHC.
Once the International Linear Collider (ILC)~\cite{ILC}
is realized, the $hhh$ coupling can be measured
with $10\%$ accuracy at the upgraded version with $\sqrt{s}=1~{\rm TeV}$
~\cite{ILCHiggsWhitePaper,Fujii:2015jha}.
The Compact LInear Collider (CLIC)~\cite{CLIC} 
is also expected to be capable of testing the $hhh$ coupling at the similar accuracy, 
while the measurement of the $hhh$ coupling is beyond the scope 
of the Future Circular Collider of electrons and positrons 
(FCC-ee)~\cite{FCC-ee}
due to the lack of its center-of-mass energy.
At future hadron colliders with the collision energy to be 100~TeV, 
the similar accuracy to the ILC may be expected for the measurement of the $hhh$ coupling as discussed in Ref.~\cite{He:2015spf}. 

An alternative method to approach EWBG is
observation of the stochastic gravitational waves (GWs)
produced by strongly first order electroweak phase transition (EWPT)~\cite{Grojean:2006bp}.
On February 11th, 2016, 
the first direct detection of GWs emitted by the merger of black holes
at Advanced LIGO was reported~\cite{Abbott:2016blz}.
It is expected that the ground-based experiments, 
KAGRA~\cite{Somiya:2011np}, Advanced LIGO~\cite{Harry:2010zz}
and Advanced VIRGO~\cite{Accadia:2009zz}
will soon
observe many GW signals from binary systems of neutron stars or black holes.
Thereby we have entered the new era of GW astronomy: 
physical phenomena can be probed through the observations of GWs.
Although the main target of the above ground-based experiments is GWs
from astronomical phenomena, 
future space-based experiments such as eLISA~\cite{Seoane:2013qna}, 
DECIGO~\cite{Kawamura:2011zz} and BBO~\cite{Corbin:2005ny}
have the sensitivity to investigate some cosmological phenomena
including strongly 1stOPTs and cosmic inflation at the early Universe.

The spectrum of stochastic GWs produced by strongly 1stOPT has been investigated in the literature.
The GW amplitude and spectrum arising from dynamics of bubbles produced 
during the phase transition have been investigated 
in Refs.~\cite{Kamionkowski:1993fg,Huber:2008hg,Espinosa:2010hh,No:2011fi,Hindmarsh:2013xza,Hindmarsh:2015qta,Caprini:2015zlo}.
Detectability of GWs from strongly 1stOPTs at GW experiments has been discussed in Refs.~\cite{Grojean:2006bp,Leitao:2012tx,Kikuta:2014eja,Jinno:2015doa,Leitao:2015fmj,Jaeckel:2016jlh,Dev:2016feu}.
EWPT and resulting GWs in models where the Higgs potential possesses
higher dimensional operators have been studied in Refs.~\cite{Delaunay:2007wb,Huang:2016odd}.
GW spectra from the strongly first order EWPT 
triggered by non-decoupling effects of new particles
have been investigated in models with extended scalar 
sectors~\cite{Espinosa:2007qk,Espinosa:2008kw,Kakizaki:2015wua},
and supersymmetric models~\cite{Apreda:2001us}.
In the case of the minimal SUSY SM (MSSM), the required strongly 1stOPT cannot be realized according to negative results of stop searches at the LHC.
Meanwhile, 1stOPT and GWs triggered by non-thermal effect at the tree level 
have been analyzed in the model with a singlet scalar 
extension~\cite{Ashoorioon:2009nf}, 
the left-right symmetric model~\cite{Sagunski:2012ufa}, 
and the Next-to-MSSM (NMSSM)~\cite{Apreda:2001us,Huber:2015znp}.
GWs in the scenario where EWBG is viable 
have been investigated in Ref.~\cite{No:2011fi}. 

In this paper, we discuss complementarity between 
measurements of the $hhh$ coupling and the spectrum of GWs 
in testing the strongly 1stOPT of the electroweak symmetry in models with additional isospin singlet scalar fields 
with and without CSI. 
We demonstrate that
the combination of the measurements of the $hhh$ coupling 
at future electron-positron colliders
and GWs at the space-based interferometers can probe physics behind the EWSB. 
As an example, we consider a set of viable CSI models with additional scalar fields 
where the EWSB is directly caused by thermal loop effects.

The idea of CSI has been originally introduced by Bardeen 
as a paradigm to avoid the hierarchy problem~\cite{Bardeen}.
In the CSI models,
utilizing the mechanism by Coleman and Weinberg~\cite{Coleman:1973jx},
a mass scale is generated through the dimensional transmutation
although any parameter with mass dimension is not introduced
in the original Lagrangian.
Based on CSI, many phenomenological models for EWSB have been proposed
and investigated~\cite{Funakubo:1993jg,Fuyuto:2015vna,Karam:2015jta,Gildener:1976ih,Takenaga:1993ux,Lee:2012jn,Endo:2015ifa,Endo:2015nba,Plascencia:2015xwa,Hashino:2015nxa,Endo:2016koi}.
Such a class of CSI models can survive current experimental tests, and have the following unique features:
a general upper bound on the lightest scalar boson is obtained;
the deviation in the Higgs-photon-photon coupling is almost fixed by
the number of charged scalar bosons; and 
the deviation in the $hhh$ coupling is universally 
as large as $70\%$~\cite{Hashino:2015nxa}.
These CSI models also predict strongly 1stOPT for EWPT. 

In Ref.~\cite{Kakizaki:2015wua},
it is shown that the spectrum of GWs from the EWPT 
strongly depends on the number of extra fields and their masses, and
is useful in revealing the Higgs potential and underlying physics.
We emphasize that the combination of future measurements
of the $hhh$ coupling and stochastic GW signals
would be a powerful tool in distinguishing the extended Higgs sectors.
For concreteness, we focus on the CSI models
where $N$ extra isospin singlet scalars obey
a global $O(N)$ symmetry, 
and discuss how these models can be differentiated
from similar extended models 
by means of the measurements of the $hhh$ coupling and the GW spectrum.
We find that they are synergetic in identifying specific models of 
EWSB in more details. 

This paper is organized as follows.
In Sec.~\ref{subsec:massless}, we briefly review the CSI $O(N)$ model.
The dynamics of 1stOPT is discussed in Sec.~\ref{sec:1stOPT}. 
In Sec.~\ref{sec:GW}, we discuss detectability of 
GW signals in the CSI models based on new fitting functions provided in Ref.~\cite{Caprini:2015zlo}.
We also compare the results in the CSI models to those 
in the models equipped with mass parameters.
Sec.~\ref{sec:conclusion} is devoted to conclusion.
The Landau pole in the CSI $O(N)$ model is discussed in Appendix~\ref{sec:landaupole}, 
and GW signals evaluated by using the conventional fitting functions given in
Ref.~\cite{Grojean:2006bp} are summarized in Appendix~\ref{sec:comparison}.

\section{Scale invariant $O(N)$ models}
\label{subsec:massless}

In this paper, we consider the following two types of $O(N)$ scalar models 
as viable examples of strongly 1stOPT, and compare both of them. 
One is the models where additional isospin singlet scalar fields are 
added to the SM Higgs sector with the mass square term which is 
supposed to be negative for EWSB. 
The $hhh$ coupling and the spectrum of GWs have been discussed in these models in Ref.~\cite{Kakizaki:2015wua}. 
Another type is the similar models but with CSI. 
In this section, we briefly review important phenomenological 
aspects of the $O(N)$ scalar model with CSI according to the work of Ref.~\cite{Hashino:2015nxa}.

The scalar sector consists of the Higgs doublet $\Phi$ and $N$ additional isospin singlet scalars $\vec{S}=(S_1^{},S_2^{}, ... , S_N^{})^T$~\cite{Endo:2015ifa,Hashino:2015nxa}.
Since the dynamics of the EWPT is basically controlled by
the mass scale of additional particles and 
the number of these extra degrees of freedom,
our results obtained are applicable to a wide class of CSI models for EWSB.
Imposing the CSI, the scalar potential is given by
\footnote{
  As far as the $O(N)$ symmetry of the CSI models is exact,
  the extra scalars are stabilized and contribute to
  the relic abundance of the dark matter.
  Applying the standard thermal dark matter 
  production scenario to these CSI models,
  the cases with $N > 4$ are not favored in the light
  of dark matter direct detection experiments~\cite{Endo:2015nba}.
  Investigating phenomenology of dark matter 
  is beyond the scope of this paper.
}
\begin{align}
  V_0(\Phi, \vec{S})=
  \frac{\lambda}{2}|\Phi|^4
  +\frac{\lambda_S}{4}|\vec{S}|^4
  +\frac{\lambda_{\Phi S}}{2}|\Phi|^2|\vec{S}|^2.
\end{align}
As far as the CSI is maintained,
the Vacuum Expectation Value (VEV) of the Higgs doublet
as well as the mass of the Higgs boson vanishes.
The CSI is violated at the quantum level, 
and a mass scale is introduced 
due to the mechanism by Coleman and Weinberg~\cite{Coleman:1973jx}.

The vacuum structure is investigated along the flat direction
of the tree-level potential
using the method developed by Gildener and Weinberg~\cite{Gildener:1976ih}.
Introducing the order parameter $\varphi$ along the flat direction as
\begin{align}
  \left\langle \Phi \right\rangle=
  \left(
  \begin{array}{c}
    0 \\
    \frac{1}{\sqrt{2}}\varphi
  \end{array}
  \right) ,
\end{align}
the one-loop effective potential at zero temperature is 
\begin{align}
  V_1(\varphi) 
  = \sum_i \frac{n_i}{64\pi^2} M_i^4(\varphi) 
  \left( \ln \frac{M_i^2(\varphi)}{Q^2} - c_i \right)\ , 
  \label{eq:v1}
\end{align}
where $Q$ is the renormalization scale.
Here, $M_i(\varphi)$ and $n_i$ are the field-dependent masses
and the degrees of freedom of particles $i$, respectively.
The degrees of freedom are given by
\begin{align}
  &n_{W_T^\pm}=4, \quad n_{W_L^\pm}=2, \quad n_{Z_T}=2, \quad n_{Z_L}=1, \nonumber \\
  &n_{\gamma_T}=2, \quad n_{\gamma_L}=1, \quad n_{t}= n_{b}=-12, \quad n_{S}=N.
\end{align}
We take the $\overline{\rm DR}$ scheme~\cite{DRbar}, where $c_i=3/2$ irrespective
of particle spins.
The one-loop contribution from the Higgs boson is absent
as this effect arises from higher-order corrections.

In the CSI models, the field-dependent mass squared of 
the field $i$ is proportional to 
$\varphi^2$ 
as $M_i^2(\varphi)= m_i^2 \varphi^2/v^2$.
The effective potential is rewritten as
\begin{align}
  V_{\eff}(\varphi, T=0)
  = A\varphi^4+B\varphi^4 \ln\frac{\varphi^2}{Q^2}, 
\end{align}
where
\begin{align}
  A
  = \sum_{i=W^\pm, Z, \gamma, t, b, S} 
  \frac{n_i}{64\pi^2 v^4} m_i^4 \left(\ln \frac{m_i^2}{v^2}-c_i\right), \quad
  B
  = \sum_{i=W^\pm, Z, \gamma, t, b, S} \frac{n_i}{64\pi^2 v^4} m_i^4. 
\end{align}
Using the stationary condition, we have 
\begin{align}
  \left. \frac{\partial V_{\eff}(\varphi, T=0)}{\partial \varphi}
  \right|_{\varphi=v}
  = \ln\frac{v^2}{Q^2}+\frac{1}{2}+\frac{A}{B} = 0 ,
  \label{eq:Q}
\end{align}
where the renormalization scale $Q$ is fixed. 

The mass squared of the discovered Higgs boson is obtained as
\begin{align}
  m_h^2 = \left. \frac{\partial^2V_{\eff}(\varphi, T=0)}{\partial \varphi^2}
  \right|_{\varphi=v}
  = 8 B v^2. 
\end{align}
Then, the one-loop effective potential is given in terms of 
the renormalized mass of the Higgs boson as
\begin{align}
  V_{\eff}(\varphi, T=0)
  =\frac{m_h^2}{8v^2}\varphi^4
  \left(\ln\frac{\varphi^2}{v^2}-\frac{1}{2}\right).
\end{align}
The renormalized $hhh$ coupling is obtained as
\begin{align}
  \lambda_{hhh} =
  \left. \frac{\partial^3V_{\eff}(\varphi, T=0)}{\partial \varphi^3}
  \right|_{\varphi=v}
  =40 v B 
  =\frac{5m_h^2}{v}=\frac{5}{3}\lambda_{hhh}^{\rm SM(tree)}, 
\end{align}
where $\lambda_{hhh}^{\rm SM(tree)}=3m_h^2/v$ is 
the tree-level $hhh$ coupling in the SM.
Then, the deviation in the $hhh$ coupling is defined by
\footnote{
  Due to nontrivial momentum dependence in the $hhh$
  coupling, there is a case where
  the deficit caused by the top loop contribution 
  obtained through the effective potential method is 
  compensated~\cite{Kanemura:2002vm,Kanemura:2004mg}.
  Accordingly, we take the tree level value for the SM prediction of
  the $hhh$ coupling.
}
\begin{align}
  \frac{\Delta \lambda_{hhh}}{\lambda_{hhh}^{\rm SM(tree)}}
  = \frac{\lambda_{hhh}}{\lambda_{hhh}^{\rm SM(tree)}}-1
  =\frac{2}{3}. 
\end{align}
Independently of the details of the models, the CSI models 
always predict
$\Delta \lambda_{hhh}^{}/\lambda_{hhh}^{\rm SM(tree)}\simeq 70\%$.

Such a characteristic feature of the CSI models for the deviation in the $hhh$ coupling is expected to be tested at future collider experiments by measuring the double Higgs boson production processes. 
At the high luminosity LHC with the integrated luminosity of $3000~{\rm fb}^{-1}$, the production cross section can be measured with $54\%$~\cite{LHChhh}, but a measurement of the $hhh$ coupling with the desired accuracy would be challenging. 
At the ILC with the center-of-mass energy of $\sqrt{s}=500~\GeV$ 
and the luminosity of
$L=4000~{\rm fb}^{-1}$, 
the $hhh$ coupling is expected to be determined at a precision of $27\%$~\cite{Fujii:2015jha}.
The expected accuracy on the $hhh$ coupling at the ILC stage
with $\sqrt{s}=1~{\rm TeV}$ and
$L=2000~{\rm fb}^{-1}$ ($L=5000~{\rm fb}^{-1}$) is
$16\%$ ($10\%$)~\cite{Fujii:2015jha}.
Therefore, this class of models can be tested at the ILC (and also the CLIC)
\footnote{
A future hadron collider with the collision energy of 100~TeV may also be able to measure the $hhh$ coupling with the similar accuracy~\cite{He:2015spf}. 
}. 

\section{Electroweak Phase Transition}
\label{sec:1stOPT}

In the CSI $O(N)$ models, the effective potential at
finite temperatures is given at the one-loop level by~\cite{Dolan:1973qd}
\begin{align}
  V_{\eff}(\varphi,T)= 
  \sum_{i=W^\pm, Z, \gamma, t, b, \vec{S}} \frac{n_i}{64\pi^2} 
  M_i^4(\varphi, T)  \left(\ln \frac{M_i^2(\varphi, T)}{Q^2}-c_i\right)
  + \Delta V_T(\varphi, T). 
  \label{Eq.effpot_massless}
\end{align}
The finite-temperature contribution to the effective potential is written as
\begin{align}
  \Delta V_T(\varphi,T) 
  = \frac{T^4}{2\pi^2} 
  \left\{ \sum_{i=W^\pm, Z, \gamma, \vec{S}} 
  n_i I_{\rm B}(a_i^2)+ \sum_{i=t,b} n_i I_{\rm F}(a_i^2) \right\} ,
\end{align}
where 
\begin{align}
  I_{\rm B,F}(a_i^2) = \int_0^\infty d x x^2\ln 
  \left[ 1\mp \exp \left( -\sqrt{x^2+a_i^2}\right) \right], 
  \quad a_i=M_i(\varphi, T)/T.  
\end{align}
Here, we employ a ring-improved effective potential that is obtained
by replacing the field-dependent masses in Eq.~(\ref{eq:v1}) 
by
\begin{align}
  M_i^2(\varphi)\to
  M_i^2(\varphi, T)= M_i^2(\varphi)+\Pi_{i}(T),
\end{align}
where $\Pi_{i}(T)$ is the finite temperature contribution to the 
self-energies~\cite{Carrington:1991hz}. 
The thermally corrected field-dependent masses of the electroweak 
gauge bosons are
\begin{align}
  M^{2(L, T)}_{g}(\varphi, T)
  =
  \frac{\varphi^2}{4}\begin{pmatrix} g^2&&& 
    \\ &g^2&& \\ &&g^2&gg' \\ &&gg'&g'^2
  \end{pmatrix} 
  + a^{L, T}_g T^2 \begin{pmatrix} 
  g^2&&& \\ &g^2&& \\ &&g^2& \\ &&&g'^2 \end{pmatrix} 
  \label{eq:gaugethermal}
\end{align}
in the $(W^+, W^-, W^3, B)$ basis with $a_{g}^L=11/6$, $a_{g}^T=0$.
Notice that only the self-energy for 
the longitudinal modes of the gauge bosons receive thermal corrections.
The field-dependent masses of fermions do not receive thermal corrections,
\begin{align}
  M^2_{t, b}(\varphi) = m_{t, b}^2\frac{\varphi^2}{v^2}.
\end{align}
The thermally corrected field-dependent mass of 
the singlet scalars are explained by
\begin{align}
  M_S^2(\varphi, T)=m_S^2 \frac{\varphi^2}{v^2}+\Pi_S(T), 
\end{align}
where 
\begin{align}
  \Pi_S(T) =\frac{T^2}{12v^2}[(N+2)\lambda_S^{} v^2+4m_S^2].
\end{align}

In order to generate BAU, Sakharov specified three necessary 
conditions~\cite{Sakharov:1967dj}.
One of them is to require the departure from thermal equilibrium,
which is accomplished 
if the baryon number changing sphaleron interaction quickly decouples
after the EWSB.
This criterion for the sphaleron decoupling is described by
\begin{align}
  \Gamma(T)\lesssim H(T) ,  
\end{align}
where $\Gamma(T)$ is the sphaleron interaction rate
and $H(T)$ is the Hubble parameter at $T$.
This condition requires that the EWSB must be of strongly 
first order~\cite{Kuzmin:1985mm},
\begin{align}
  \frac{\varphi_c}{T_c} \gtrsim 1 ,
  \label{ft}
\end{align}
where $\varphi_c$ is the VEV for the broken phase minimum
at the critical temperature $T_c$.
In this paper, we numerically compute $\varphi_c$ and $T_c$ 
instead of using high temperature expansion.

\begin{figure}[t]
  \centerline{\includegraphics[width=0.48\textwidth]{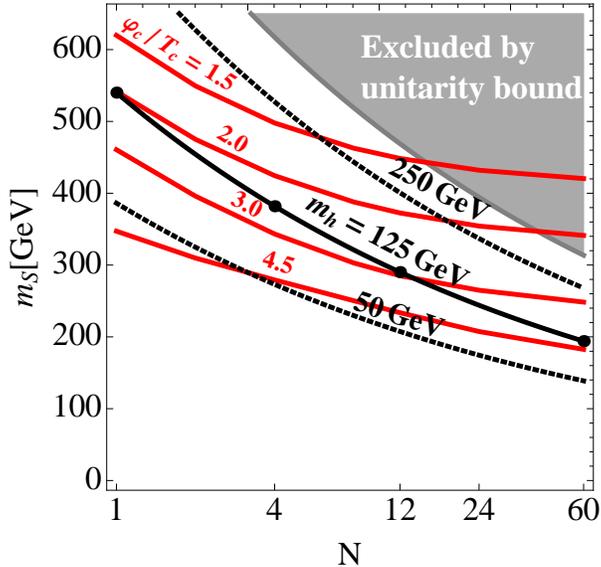}}
  \caption{The contours of the mass of the Higgs boson $m_h$ and $\varphi_c/T_c$
    in the $(N, m_S)$ plane in the CSI $O(N)$ model.}
  \label{fig:PT_masslessO(N)}
\end{figure}

In Fig.~\ref{fig:PT_masslessO(N)}, we show contours
of the mass of the Higgs boson $m_h$ and $\varphi_c/T_c$ 
in the $(N, m_S)$ plane in the CSI $O(N)$ model.
With the mass of the Higgs boson being 125~GeV,
strongly 1stOPTs of $\varphi_c/T_c>2$ are realized for $N \geq 1$.
It is characteristic of the CSI models to satisfy the condition for strongly 1stOPT for the EWSB.
The right-top shaded region is excluded by the perturbative unitarity bound~\cite{Kakizaki:2015wua}.
In order to keep perturbative unitarity, we impose $|a_i|<1/2$ 
for the eigenvalues of $S$-wave
amplitudes of two body elastic scatterings among longitudinally polarized
weak bosons and scalar bosons.
We find that it is sufficient to impose the following inequality:
\begin{align}
  \frac{1}{32\pi}
  \left[3\lambda +(N+2)\lambda_S 
  +\sqrt{\{3\lambda-(N+2)\lambda_S\}^2+4N\lambda_{\Phi S}^2}\right]<\frac{1}{2}.
  \label{pu}
\end{align}
In Fig.~\ref{fig:PT_masslessO(N)}, vanishingly small
$\lambda$ and $\lambda_S^{}$ are assumed as these parameters do not 
affect the computation of the GW spectrum discussed in the next section.

\section{Gravitational waves from strongly first order phase transition in the $O(N)$ models with and without CSI}
\label{sec:GW}

\subsection{Bubble dynamics parameters $\alpha$ and $\beta$}
\label{subsec:parameters}

We here introduce two important quantities $\alpha$ and $\beta$
which parametrize the spectrum of GWs following Ref.~\cite{Grojean:2006bp}. 
One of the key quantities that describe dynamics of vacuum bubble 
is the bubble nucleation rate per unit volume per unit time,
\begin{align}
  \Gamma(t)=\Gamma_0(t)\exp[-S_E(t)] ,
\end{align}
where $S_E$ is the Euclidean action of a critical bubble
and $\Gamma_0 \sim T^4$.
To parametrize the rate of variation of the bubble nucleation rate
at the time of the phase transition $t_t$,
we introduce the quantity $\beta$ defined as 
\begin{align}
  \beta = - \left. \frac{d S_E}{d t}\right|_{t=t_t}
  \simeq \left. \frac{1}{\Gamma}\frac{d \Gamma}{d t}\right|_{t=t_t}.
\end{align} 
The inverse of the parameter $\beta$ describes approximately the time duration
of the phase transition.
At finite temperatures, the $O(3)$-symmetric field configuration $\varphi_b^{}$ 
contributes to
the action as $S_E(T) = S_3(T)/T$,
where $S_3$ is the three dimensional Euclidean action, 
\begin{align}
  S_3 = \int d^3r \left[\frac{1}{2}  (\vec{\nabla}\varphi_b)^2
  + V_{\rm eff}^{}(\varphi_b, T)\right].
\end{align}
The bounce solution $\varphi_b$ is obtained by solving the differential equation,
\begin{align}
  \frac{d^2 \varphi_b^{}}{d r^2}
  + \frac{2}{r} \frac{d \varphi_b^{}}{d r} 
  - \frac{\partial V_{\rm eff}^{}}{\partial \varphi_b^{}} 
  =0 ,
\end{align}
with the boundary conditions
\begin{align}
  \left. \frac{d \varphi_b^{}}{d r} \right|_{r=0} = 0,
  \quad
  \lim_{r \to \infty} \varphi_b = 0.
\end{align}
It is convenient to introduce the dimensionless parameter
\begin{align}
  \tilde{\beta} = 
  \frac{\beta}{H_t} =
  T_t\frac{d}{d T}\left(\frac{S_3(T)}{T}\right)\Bigg|_{T=T_t}. 
\end{align}
In order for the nucleated vacuum bubbles to percolate through the Universe,
the nucleation rate per Hubble volume per Hubble time should reach the unity
\begin{align}
  \left.\frac{\Gamma}{H^4}\right|_{T=T_t} \simeq 1.
  \label{eq:PT}
\end{align}
This condition is converted to the relation
\begin{align}
  \frac{S_3(T_t^{})}{T_t^{}} =4\ln (T_t/H_t) 
  \simeq 140-150.   
\end{align}

Another important quantity is the false-vacuum energy density available for the evolution of bubbles
\begin{align}
  \epsilon(T)
  = - V_{\eff}(\varphi_B(T),T)+T
  \frac{\partial V_{\eff}(\varphi_B(T), T)}{\partial T},
\end{align}
where $\varphi_B(T)$ is the VEV for the broken phase minimum at $T$,
and we reset the symmetric phase minimum to zero.  We introduce the
parameter $\alpha$ parametrizing the ratio of the false-vacuum energy
density $\epsilon(T_t^{})$ to the thermal energy density
$\rho_{\rm rad}^{}(T_t^{})$ in the symmetric phase at the transition
temperature $T_t$,
\begin{align}
  \alpha = \frac{\epsilon(T_t)}{\rho_{\rm rad}(T_t)} ,
\end{align}
where the radiation energy density is given by
$\rho_{\rm rad}^{}(T) =(\pi^2/30) g_*^{}(T)T^4$, 
with $g_*^{}$ being the relativistic degrees of 
freedom in the thermal plasma.

The predictions for $\alpha$, $\tilde{\beta}$, $m_S$, $\varphi_c/T_c$,
$T_c$ and $T_t$ 
in the CSI $O(N)$ models are shown
in the top frame of Table~\ref{tab:ON}.
In the middle and bottom frames of Table~\ref{tab:ON}, 
for comparison, 
we also listed the predictions of the $O(N)$ models
with mass terms that produce
$\Delta\lambda_{hhh}^{}/\lambda_{hhh}^{\SM}=2/3 (\simeq 70\%)$ for $\sqrt{\mu_S^2}=0~\GeV$ and $\sqrt{\mu_S^2}=100~\GeV$, respectively. 
(For the details of the $O(N)$ models with mass terms; i.e. the $O(N)$ models without CSI, 
see Ref.~\cite{Kakizaki:2015wua}). 
In the $O(N)$ models without CSI, $\varphi_c/T_c$ and $\alpha$ ($\tilde{\beta}$) decrease (increases) with increasing $\sqrt{\mu_S^2}$ for a fixed value of $\Delta\lambda_{hhh}^{}/\lambda_{hhh}^{\SM}$.

\begin{figure}[t]
  \centerline{\includegraphics[width=0.48\textwidth]{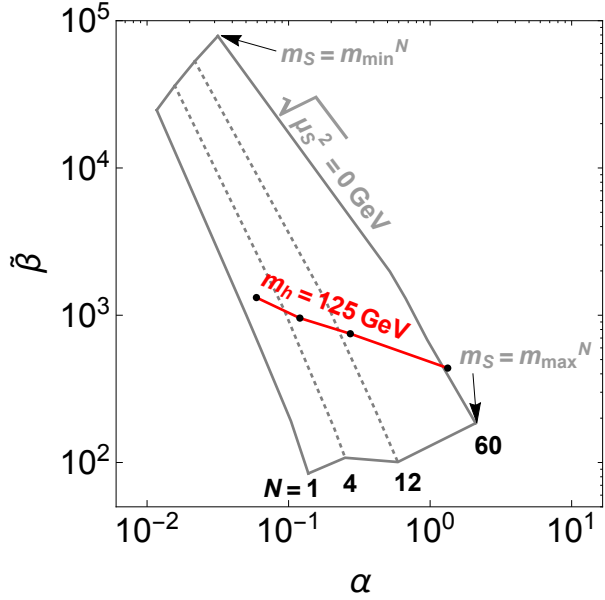}}
  \caption{Predicted values of $\alpha$ and $\tilde{\beta}$
    in the CSI $O(N)$ models (red) and $O(N)$ models without CSI for $\sqrt{\mu_S^2}=0~\GeV$ (gray).}
  \label{fig:alphabeta}
\end{figure}

In Fig.~\ref{fig:alphabeta}, the predicted values of $\alpha$ and $\beta$
in the CSI $O(N)$ models (red) and $O(N)$ models without CSI for $\sqrt{\mu_S^2}=0~\GeV$ (gray)
are plotted.
Due to the finite-temperature effect, the CSI is violated.
As a result, the predicted values of $\alpha$ and $\beta$ 
as well as $\varphi_c/T_c$ significantly change depending on  the number of the extra scalar bosons although the $hhh$ coupling is common in the CSI models.
The upper (lower) bound on $\beta$ for each $O(N)$ model without CSI 
is derived by the condition of $\varphi_c/T_c=1$ ($\Gamma/H^4|_{T=T_t^{}}=1$).
The predicted values of 
$\alpha$ and $\beta$ strongly depend
on the number of extra scalars $N$ and their mass $m_S^{}$ as
pointed out in Ref.~\cite{Kakizaki:2015wua}.
This fact opens new possibilities for resolving these models
through observation of GW signals.

\begin{table}[t]
  \footnotesize{
    \begin{flushleft}
      \begin{tabular}{|c||c|c|c|c|} \hline
$N$&1&4&12&60 \\\hline\hline
$m_S$& 
540\,GeV& 
382\,GeV&
290\,GeV&
194\,GeV \\ \hline
$\varphi_c/T_c$, $T_c$&
2.01, 102\,GeV&
2.40, 90.1\,GeV&
2.91, 76.8\,GeV&
4.11, 56.1\,GeV \\ \hline
$(\alpha, \tilde{\beta})$, $T_t$&
(0.0593, 1320), 88.5\,GeV&
(0.120, 956), 74.3\,GeV&
(0.273, 705), 59.7\,GeV&
(1.33, 438), 38.4\,GeV \\ \hline
      \end{tabular}
    \end{flushleft}
    \begin{flushleft}
      \begin{tabular}{|c||c|c|c|c|} \hline
($\sqrt{\mu_S^2}$, $N$)&(0\,GeV, 1)&(0\,GeV, 4)&(0\,GeV, 12)&(0\,GeV, 60) \\ \hline\hline
$m_S$&
510\,GeV&
361\,GeV&
274\,GeV&
183\,GeV \\ \hline
$\varphi_c/T_c$, $T_c$&
1.62, 119\,GeV&
2.03, 102\,GeV&
2.54, 85.6\,GeV&
3.65, 61.5\,GeV \\ \hline
$(\alpha, \tilde{\beta})$, $T_t$&
(0.0303, 3320), 111\,GeV&
(0.0695, 2180), 92.5\,GeV&
(0.164, 1600), 74.8\,GeV&
(0.739, 1090), 50.3\,GeV \\ \hline
      \end{tabular}
    \end{flushleft}
    \begin{flushleft}
      \begin{tabular}{|c||c|c|c|c|} \hline
($\sqrt{\mu_S^2}$, $N$)&(100\,GeV, 1)&(100\,GeV, 4)&(100\,GeV, 12)&(100\,GeV, 60) \\\hline\hline
$m_S$&
524\,GeV&
380\,GeV&
299\,GeV&
219\,GeV \\ \hline
$\varphi_c/T_c$, $T_c$&
1.56, 121\,GeV&
1.89, 106\,GeV&
2.25, 92.1\,GeV&
2.89, 71.6\,GeV \\ \hline
$(\alpha, \tilde{\beta})$, $T_t$&
(0.0272, 4380), 115\,GeV&
(0.0552, 3480), 99.5\,GeV&
(0.111, 3210), 85.7\,GeV&
(0.334, 4082), 67.2\,GeV \\ \hline
      \end{tabular}
    \end{flushleft}
  }
  \caption{Predictions of the four benchmark points $N=1, 4, 12$ and $60$ 
    in the CSI $O(N)$ models (top).
    For comparison, the predictions of $O(N)$ models without CSI with $\Delta\lambda_{hhh}^{}/\lambda_{hhh}^{\SM}=2/3 (\simeq 70\%)$ are also shown for $\sqrt{\mu_S^2}=0~\GeV$ (middle) and $\sqrt{\mu_S^2}=100~\GeV$ (bottom). }
  \label{tab:ON}
\end{table}

\subsection{Spectrum of gravitational waves}
\label{subsec:GW}

During the 1stOPT in the early Universe, collisions of vacuum bubbles
and bulk motion of the plasma produce GWs.
There are three contributions to the stochastic GW background
from the 1stOPT; collisions of bubble walls, the compression waves in the plasma (sound waves) and magnetohydrodynamic (MHD) turbulence in the plasma.
Computation of the produced spectrum of the GWs from the 1stOPT depends on the details of dynamics of colliding bubbles, and complicated numerical simulations are required.
Here, we employ the fitting functions recently
provided in Ref.~\cite{Caprini:2015zlo}
for estimating the GW spectrum from the parameters $\alpha$ and $\beta$. 
In our previous work in the $O(N)$ models without CSI~\cite{Kakizaki:2015wua}, we employed the fitting function for GWs 
from bubble collisions in the envelope approximation in Ref.~\cite{Kamionkowski:1993fg} and that for GWs from plasma turbulence in Ref.~\cite{Nicolis:2003tg}. 
In Appendix~\ref{sec:comparison}, we discuss differences between these fitting functions and those in Ref.~\cite{Caprini:2015zlo}. 

The GW spectrum from the collision of bubble walls can 
be computed using the method referred to as the
envelope approximation~\cite{Huber:2008hg}.
A fit to the spectrum obtained by numerical simulations is given by 
\begin{align}
  \Omega_{\rm env} (f) h^2 
  =\widetilde{\Omega}_{\rm env} h^2 
  \frac{3.8 (f/\tilde{f}_{\rm env})^{2.8}}{1+2.8 (f/\tilde{f}_{\rm env})^{3.8}},  
\end{align}
where the peak of the energy density is given by
\begin{align}
  \widetilde{\Omega}_{\rm env} h^2\simeq 
  1.67 \times 10^{-5}  \times\left(\frac{0.11v_b^3}{0.42+v_b^2}\right) 
  \tilde{\beta}^{-2}
  \left(\frac{\kappa_\varphi \alpha}{1+\alpha}\right)^2 
  \left(\frac{100}{g^t_\ast}\right)^{1/3}, 
\end{align}
at the peak frequency
\begin{align}
  \tilde{f}_{\rm env} \simeq
  1.65 \times 10^{-5}~\Hz \times \left(\frac{0.62}{1.8-0.1v_b+v_b^2}\right)
  \tilde{\beta}
  \left(\frac{T_t}{100~\GeV}\right)\left(\frac{g^t_\ast}{100}\right)^{1/6}.
\end{align}
Here, $v_b$ is the wall velocity of the bubble, 
and the parameter $\kappa_\varphi$ is the ratio of the vacuum energy transferred to the bubble motion.
The relativistic degrees of freedom at the transition temperature
is denoted by $g_*^t(=g_*^{}(T_t^{}))$.

As for the contribution to GWs from sound waves, 
a fitting function to the numerical simulations is obtained as~\cite{Caprini:2015zlo}
\begin{align}
  \Omega_{\rm sw} (f) h^2 
  =\widetilde{\Omega}_{\rm sw} h^2 \times
    (f/\tilde{f}_{\rm sw})^3
  \left(\frac{7}{4+3(f/\tilde{f}_{\rm sw})^2}\right)^{7/2},  
  \end{align}
where the peak energy density is given by 
\begin{align}
  \widetilde{\Omega}_{\rm sw} h^2
  \simeq 2.65 \times 10^{-6} v_b 
  \tilde{\beta}^{-1}
  \left(\frac{\kappa_v \alpha}{1+\alpha}\right)^2
  \left(\frac{100}{g^t_\ast}\right)^{1/3}, 
\end{align}
at the peak frequency
\begin{align}
  \tilde{f}_{\rm sw} \simeq 1.9 \times 10^{-5}~\Hz \frac{1}{v_b}
  \tilde{\beta}
  \left(\frac{T_t}{100~\GeV}\right)
  \left(\frac{g^t_\ast}{100}\right)^{1/6}. 
\end{align}

In addition to the compression waves, turbulent motion in the plasma also
contributes to the GW spectrum.
Taking MHD effects on the ionized plasma into account,
the GW spectrum from the MHD turbulence 
is~\cite{Binetruy:2012ze,Caprini:2009yp}
\begin{align}
  \Omega_{\rm turb} (f) h^2 
  =\widetilde{\Omega}_{\rm turb} h^2 \times
  \frac{(f/\tilde{f}_{\rm turb})^3 }{(1+ f/\tilde{f}_{\rm turb})^{11/3} 
  (1+8 \pi f/h_t^{})},
\end{align}
where 
\begin{align}
  \widetilde{\Omega}_{\rm turb} h^2
  \simeq 3.35 \times 10^{-4} v_b 
  \tilde{\beta}^{-1}
  \left(\frac{\epsilon \kappa_v \alpha}{1+\alpha}\right)^{3/2} 
\left(\frac{100}{g^t_\ast}\right)^{1/3}, 
\end{align}
and 
\begin{align}
  \tilde{f}_{\rm turb} \simeq 2.7 \times 10^{-5}~\Hz \frac{1}{v_b}
  \tilde{\beta}
  \left(\frac{T_t}{100~\GeV}\right)
  \left(\frac{g^t_\ast}{100}\right)^{1/6},  
\end{align}
The MHD turbulent contribution depends on the Hubble parameter 
at the time of GW production, whose redshifted value is
\begin{align}
  h_t^{}=1.65 \times 10^{-5}~\Hz
  \left(\frac{T_t}{100~\GeV}\right) 
  \left(\frac{g^t_\ast}{100}\right)^{1/6}.
\end{align}
According to the recent simulations~\cite{Hindmarsh:2015qta}, 
the fraction of the turbulent bulk motion is at most $\epsilon \simeq 5-10\%$.
In our later numerical analysis, we set $\epsilon = 0.05$.

\begin{figure}[t]
  \centerline{\includegraphics[width=0.48\textwidth]{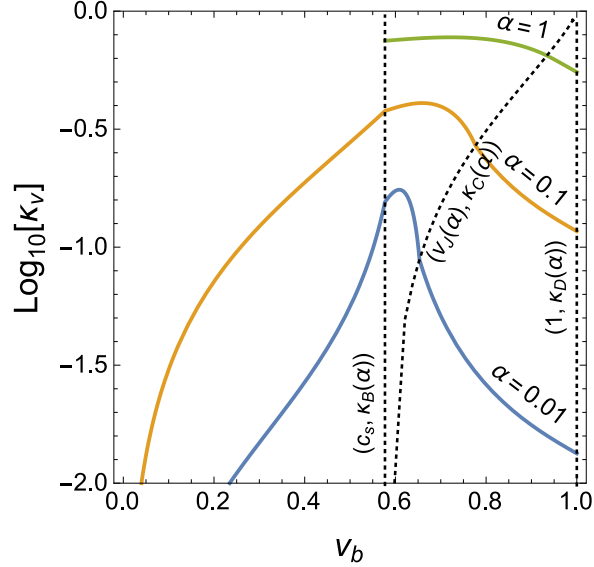}}
  \caption{
    The efficiency factor $\kappa_v^{}$ as a function of 
    $v_b^{}$ for $\alpha=0.01, 0.1$ and $1$. 
    The cases for $v_b^{}=c_s^{}$ and $v_b^{}=v_J^{}$ are
    plotted with dotted lines. 
  }
  \label{fig:kappa}
\end{figure}

The efficiency factor $\kappa_v$ denotes the fraction of the vacuum energy 
that is transformed into the bulk motion of the plasma fluid,
and given as a function of $\alpha$ and $v_b$.
We refer to results obtained in Ref.~\cite{Espinosa:2010hh},
\begin{align}
  \kappa_v(v_b, \alpha)\simeq
  \begin{cases}
    & \frac{ c_s^{11/5}\kappa_A \kappa_B }{(c_s^{11/5} -  v_b^{11/5} )\kappa_B
      +  v_b c_s^{6/5} \kappa_A}\quad (\text{for}~ v_b \lesssim c_s ) \\
    & \kappa_B + ( v_b - c_s) \delta\kappa 
    + \frac{( v_b - c_s)^3}{ (v_J - c_s)^3} [ \kappa_C - \kappa_B -(v_J - c_s) 
    \delta\kappa ]
    \quad (\text{for}~ c_s <  v_b < v_J) \\
    & \frac{ (v_J - 1)^3 v_J^{5/2}  v_b^{-5/2}
      \kappa_C \kappa_D }
    {[( v_J -1)^3 - ( v_b-1)^3] v_J^{5/2} \kappa_C
      + ( v_b - 1)^3 \kappa_D }
    \quad ( \text{for}~ v_J \lesssim v_b )
  \end{cases},    
\end{align}
where
\begin{align}
  \kappa_A 
  &\simeq v_b^{6/5} \frac{6.9 \alpha}{1.36 - 0.037 \sqrt{\alpha} + \alpha},
  \quad
  \kappa_B 
  \simeq \frac{\alpha^{2/5}}{0.017+ (0.997 + \alpha)^{2/5} },
    \nonumber \\
  \kappa_C 
  &\simeq \frac{\sqrt{\alpha}}{0.135 + \sqrt{0.98 + \alpha}},
  \quad
  \kappa_D 
  \simeq \frac{\alpha}{0.73 + 0.083 \sqrt{\alpha} + \alpha}.
\end{align}
The efficiency factor is treated separately depending on whether
the wall velocity $v_b^{}$ exceeds the velocity of sound ($c_s=0.577$) or
\begin{align}
  v_J=\frac{\sqrt{ 2/3 \alpha  + \alpha^2 } + \sqrt{1/3}}{1 + \alpha}.  
\end{align}
The derivative of $\kappa_v$ with respect to $v_b$ 
at $v_b = c_s$ is approximately given by 
\begin{align}
  \delta\kappa\simeq -0.9\ln{\frac{\sqrt{\alpha}}{1 + \sqrt{\alpha}}}.  
\end{align}
The dependence of the efficiency factor $\kappa_v$ on $v_b$ and $\alpha$ is
shown in Fig.~\ref{fig:kappa}. 
For large $\alpha$ and small $v_b$,
no consistent solution to hydrodynamic equations exists~\cite{Espinosa:2010hh,No:2011fi}. 

If the vacuum energy released at the phase transition is adequately large,
it is possible for the bubble wall velocity to approach the speed of 
light~\cite{Espinosa:2010hh,Caprini:2015zlo}.
The minimum value of $\alpha$ for realizing 
this runaway bubble wall ($v_b = 1$) is 
approximately~\cite{Espinosa:2010hh,Caprini:2015zlo}
\begin{align}
  \alpha_\infty \simeq \frac{30}{24\pi^2 g_\ast^t T_t^2} 
  \sum_i c_i \left[ M_i^2 (\varphi_t) - M_i^2 (0) \right],
\end{align}
where $c_i^{}=|n_i^{}|$ ($|n_i^{}|/2$) for bosons (fermions).
Particles that are light in the symmetric phase and 
become non-relativistic after the phase transition
contribute to the sum on the right-hand side of the above equation.
In the runaway case, 
abundant energy can be used for accelerating the bubble wall,
enhancing the contribution from the wall collision to GWs.
The efficiency factors are given by
\begin{align}
  \kappa_\varphi^{} = 1 - \frac{\alpha_\infty}{\alpha} \geq 0, \quad
  \kappa_v = \frac{\alpha_\infty}{\alpha}\kappa_\infty, 
\end{align}
where 
$\kappa_\infty^{} = \kappa_v^{}(1, \alpha_\infty^{}) = \kappa_D^{}(\alpha_\infty^{})$. 
In the non-runaway case ($v_b<1$), on the other hand, 
the wall collision contribution is small~\cite{Espinosa:2010hh}.

\subsection{Detectability of gravitational wave signals}
\label{subsec:GW_result}

We here show our numerical results of the spectrum of GWs 
from strongly 1stOPT in the $O(N)$ model with and without CSI. 
We then discuss the detectability of the GWs at the future space-based
interferometers, eLISA and DECIGO. 

\begin{figure}[t]
  \begin{minipage}[t]{0.48\hsize}
  \centerline{\includegraphics[width=1\textwidth]{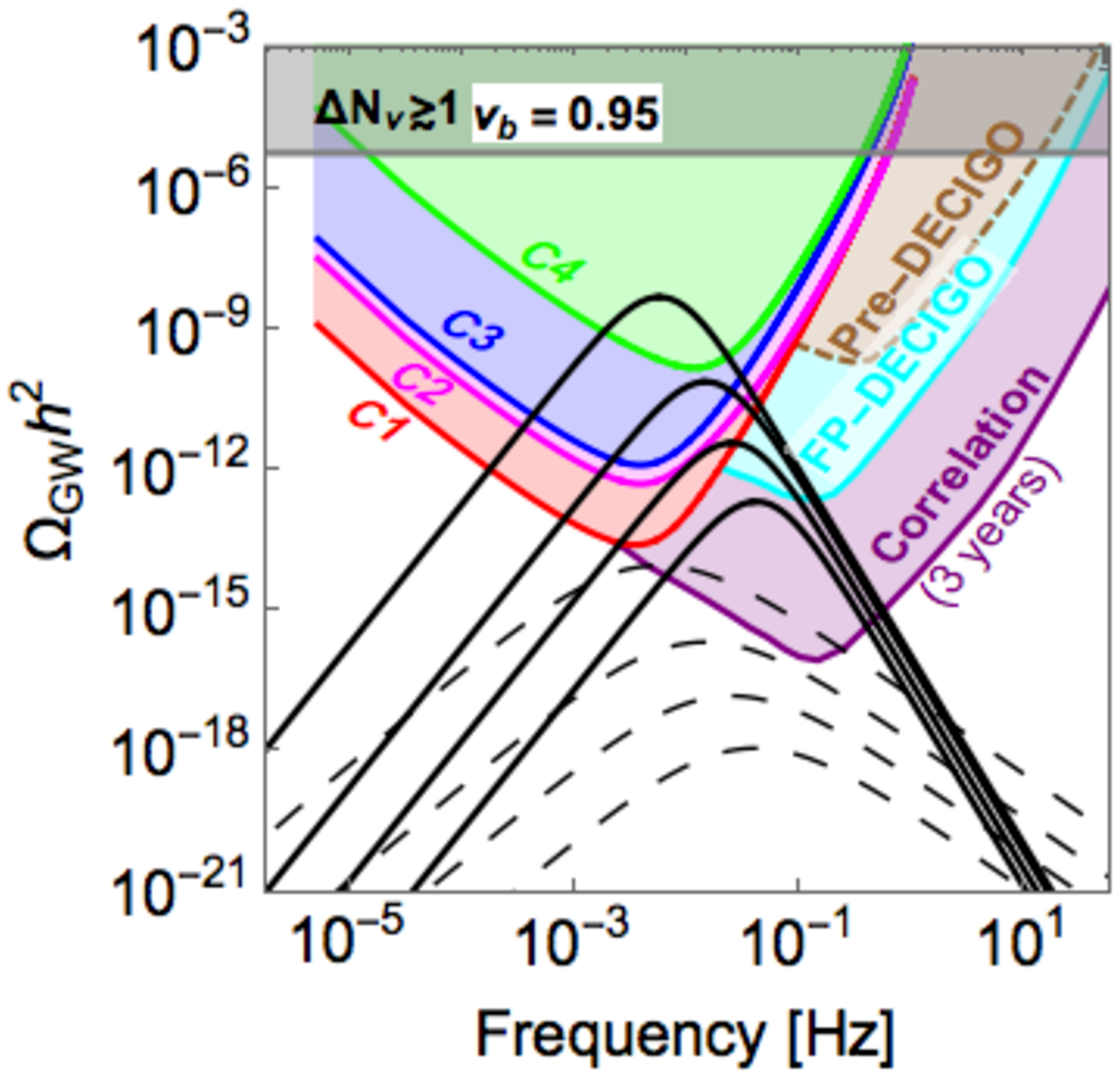}}
  \end{minipage}
  \begin{minipage}[t]{0.48\hsize}
  \centerline{\includegraphics[width=1\textwidth]{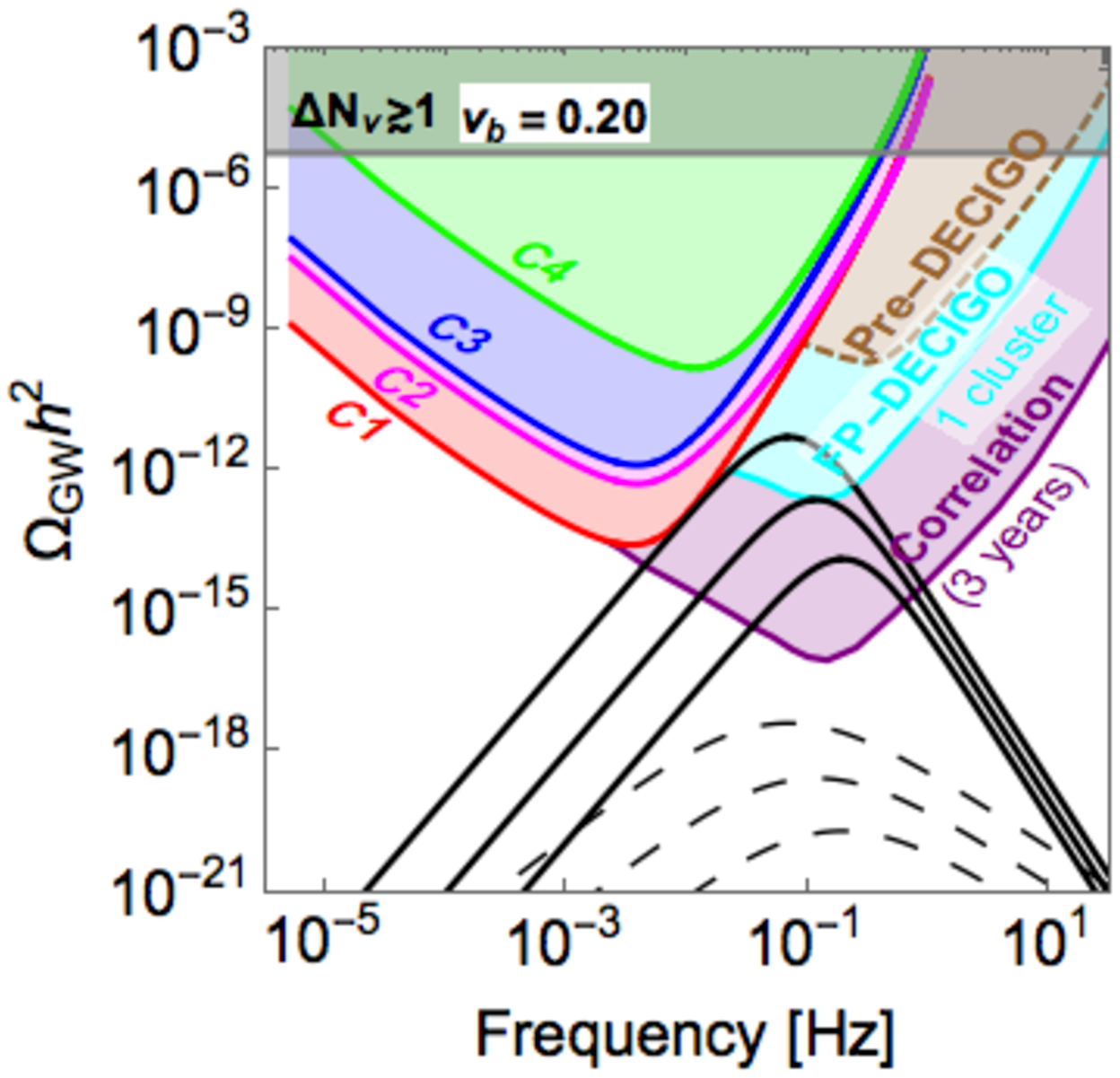}}
  \end{minipage}
  \begin{minipage}[t]{0.48\hsize}
  \centerline{\includegraphics[width=1\textwidth]{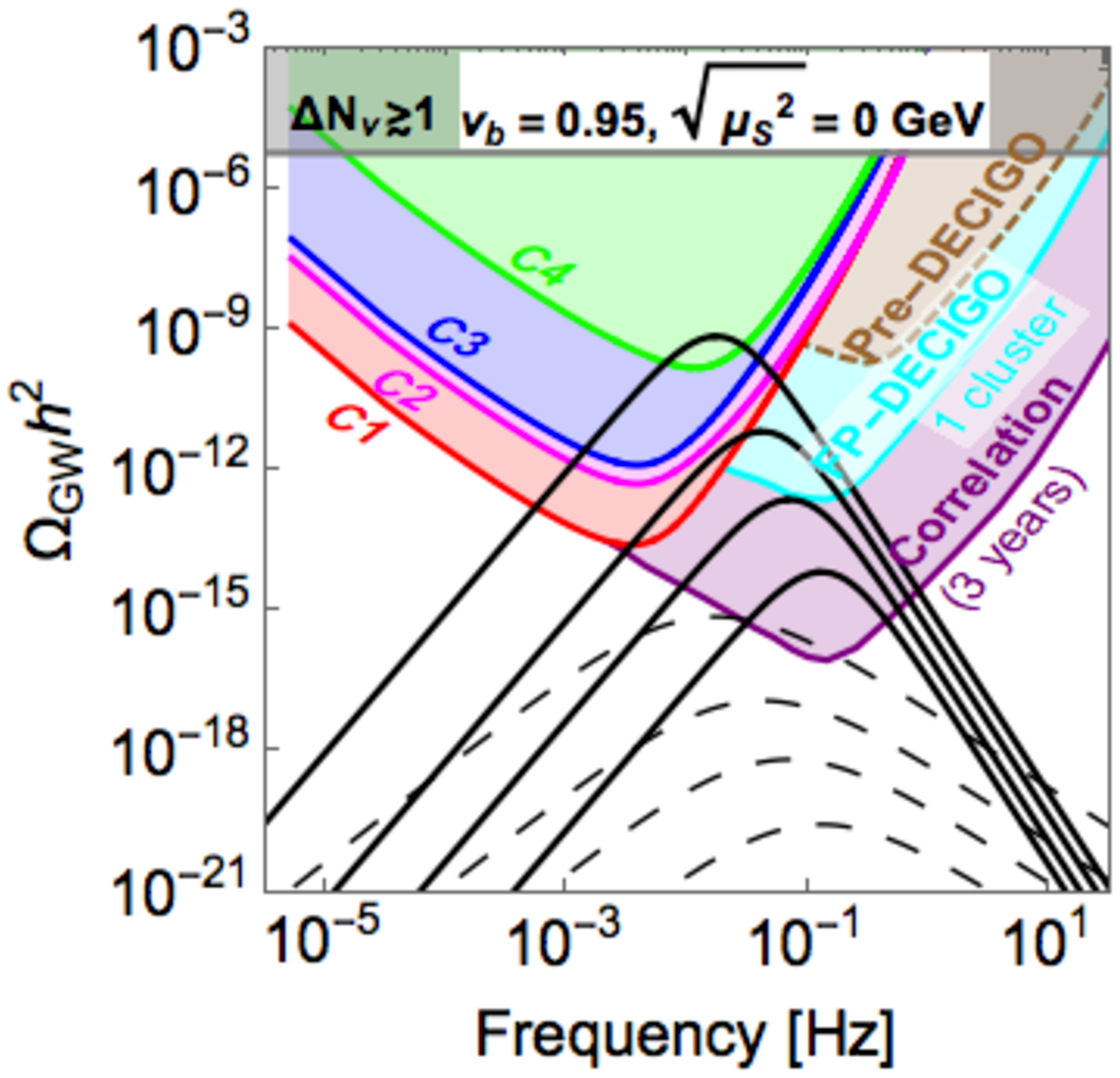}}
  \end{minipage}
  \begin{minipage}[t]{0.48\hsize}
  \centerline{\includegraphics[width=1\textwidth]{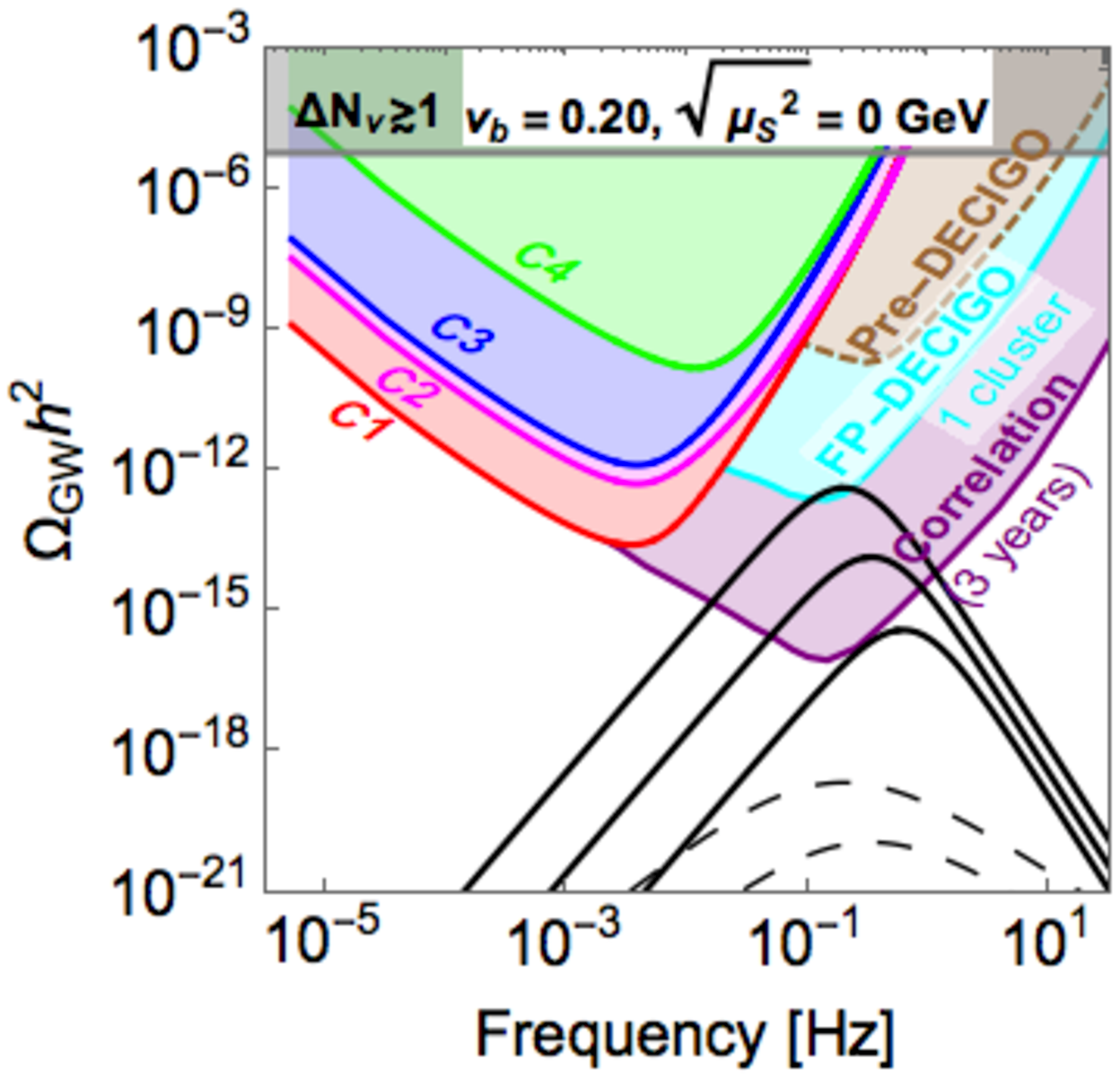}}
  \end{minipage}
  \caption{
    GW spectra in the CSI $O(N)$ models (top frames)
    and $O(N)$ models without CSI with
    $\Delta\lambda_{hhh}^{}/\lambda_{hhh}^{\SM}=2/3 (\simeq 70\%)$ (bottom frames)
    for $v_b^{}=0.95$ (left) and $v_b=0.2^{}$ (right).
    The black curves correspond to the contributions 
    from the sound waves (solid) 
    and turbulence (dashed)
    for $N=1,4,12$ and $60$ (left) for $N=1,4$ and $12$ (right) from the bottom.
  }
  \label{fig:GWspectrum}
\end{figure}

In Fig.~\ref{fig:GWspectrum} (top frames),
we show the predicted GW spectra produced from the EWPT
in the CSI $O(N)$ models using the formulae provided above.
The black curves correspond to the contributions 
from the sound waves (solid)
and turbulence (dashed)
for $N=1,4,12$ and $60$ from the bottom.
See Table~\ref{tab:ON} for the predicted values.
The CSI $O(N)$ model with $N=60$ requires $v_b^{}>0.65$ to obtain a
consistent solution to hydrodynamics equations,
and is not displayed in the right frame.
It turns out that the contribution from sound waves tends to dominate.
The colored regions show experimental sensitivities
expected at future space-based interferometers, 
eLISA~\cite{Klein:2015hvg,Caprini:2015zlo,PetiteauDataSheet} 
and DECIGO~\cite{Kawamura:2011zz}.
The labels ``C1'', ``C2'', ``C3'' and ``C4''
correspond to four different configurations
of eLISA
provided in Table 1 in Ref.~\cite{Caprini:2015zlo} whereas
the labels ``Pre-DECIGO'', ``FP-DECIGO'' and ``Correlation'' are DECIGO 
designs~\cite{Kawamura:2011zz}.
As for the wall velocity, we take two reference values $v_b^{}=0.95$ (left)
and $v_b^{}=0.2$ (right).
For $v_b^{}=0.95$,
even the eLISA C4 design has sensitivity to observe the predicted GW spectra
in the CSI model with $N=60$.
The eLISA C1 design can probe the CSI models with $N \gtrsim 4$.
The energy density of GWs is suppressed
and its peak frequency is enhanced for $v_b^{}=0.2$.
In this case, 
DECIGO will be sensitive even to the model with $N = 1$ taking the correlation of nearby two clusters.
For smaller values of $v_b^{}$, the generated GWs are too small
to detect at the eLISA C1 design.
For successful EWBG, however, subsonic wall velocities as small as 
$v_b<0.15-0.3$ are generically required~\cite{Joyce:1994fu}.
Contrary to this argument, Ref.~\cite{No:2011fi} points out
there are some cases where EWBG is possible even if
wall velocities are supersonic and GW signals are large.
The upper bound on the GWs is 
derived from non-observation of extra 
radiation~\cite{planck, Agashe:2014kda}, and
indicated by $\Delta N_\nu \gtrsim 1$.
We comment on foreground noise that may obscure the GW signals from the EWPT.
The maximum value of the contribution from extragalactic white dwarf binaries
is estimated to be $\Omega h^2 \sim 10^{-11}$ at 
$f \sim 10^{-2}~{\rm Hz}$~\cite{Schneider:2010ks},
while that from galactic white dwarf binaries is significant
for $f<10^{-3}$ Hz~\cite{Nelemans:2009hy}.
Therefore, the sound wave contribution to GWs spectrum is not screened
by these astrophysical sources for large $N$ and large wall velocities.
For comparison, the GW spectra predicted in
the $O(N)$ models without CSI that coincidentally leads to
$\Delta\lambda_{hhh}^{}/\lambda_{hhh}^{\SM}=2/3 (\simeq 70\%)$
are shown in Fig.~\ref{fig:GWspectrum} (bottom frames). 
The $O(N)$ model without CSI with $N=60$ requires $v_b>0.49$, and is not 
displayed in the right frame.

\begin{figure}[t]
  \centerline{\includegraphics[width=0.48\textwidth]{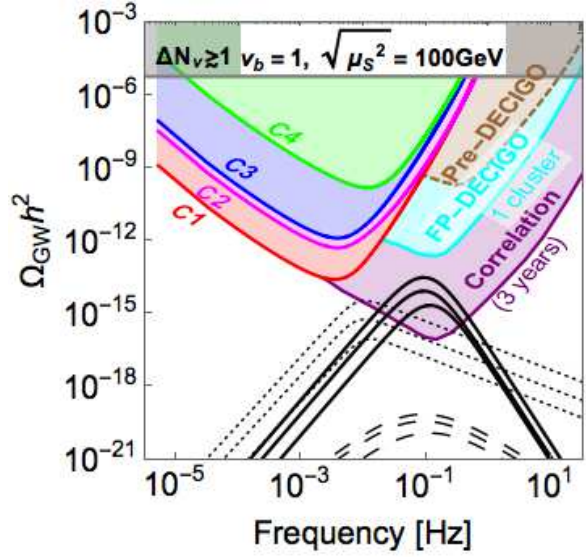}}
  \caption{
    GW spectra in the $O(N)$ models without CSI with
    $\sqrt{\mu_S^2}=100~\GeV$ and 
    $\Delta\lambda_{hhh}^{}/\lambda_{hhh}^{\SM}=2/3 (\simeq 70\%)$
    for the runaway case ($v_b^{}=1$).
    The black curves correspond to the contributions 
    from the sound waves (solid), bubble collision (dotted) 
    and turbulence (dashed) for $N=4,12$ and $60$ from the bottom.
  }
  \label{fig:GWspectrum2}
\end{figure}

The GW spectra for the runaway case ($v_b^{}=1$) are shown in 
Fig.~\ref{fig:GWspectrum2}. 
The runaway bubble wall can be realized for
$N=4, 12$ and $60$ in the $O(N)$ models without CSI
with $\sqrt{\mu_S^2}=100~\GeV$ among our benchmark points.
In the CSI $O(N)$ models and other $O(N)$ models without CSI,
a large number of singlet scalars become heavy during the phase transition
and contribute to the friction on the bubble wall, preventing the wall from the 
runaway behavior.

\begin{figure}[t]
  \begin{minipage}[t]{0.48\hsize}
 \centerline{\includegraphics[width=1\textwidth]{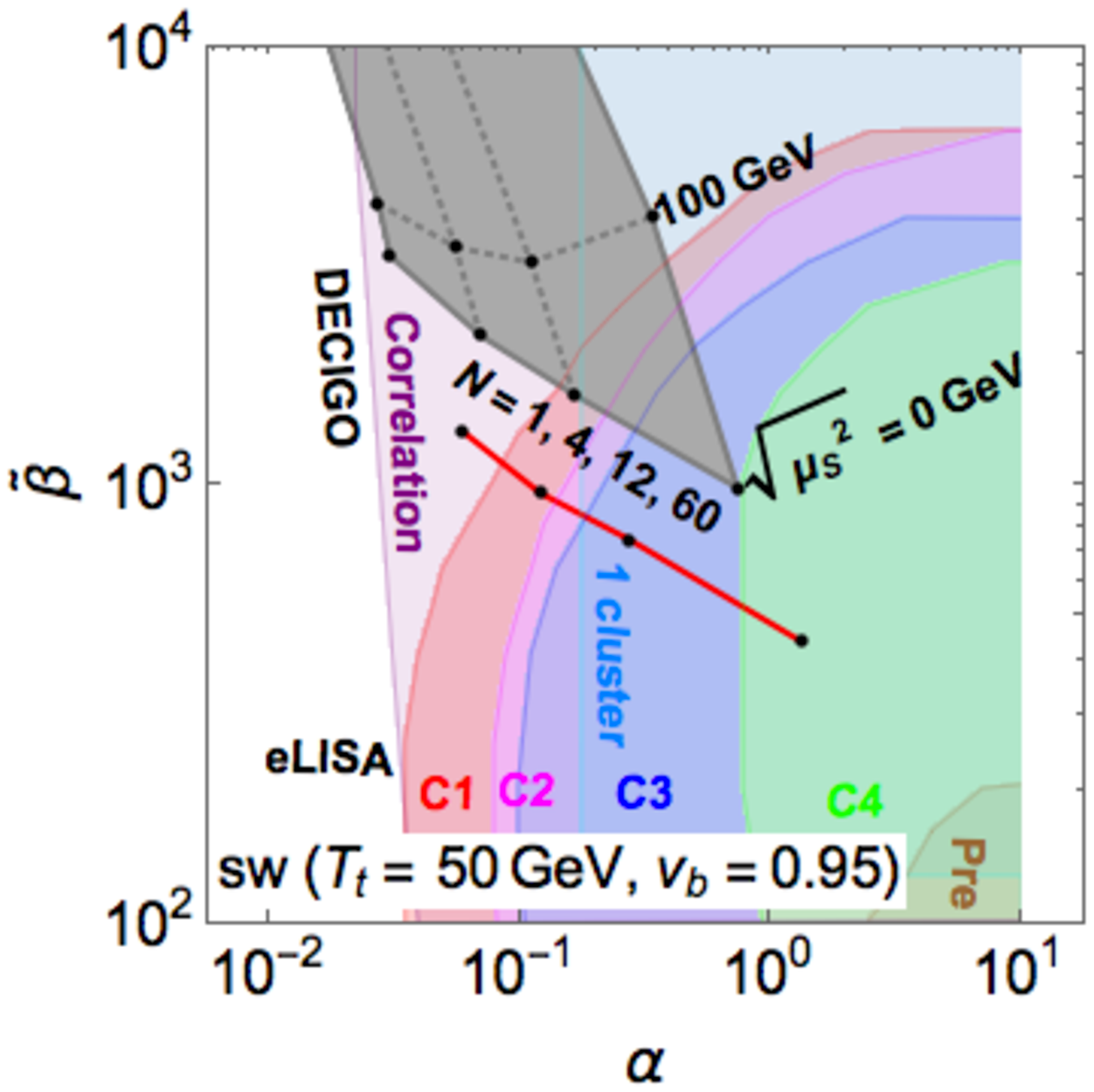}}
  \end{minipage}
  \begin{minipage}[t]{0.48\hsize}
  \centerline{\includegraphics[width=1\textwidth]{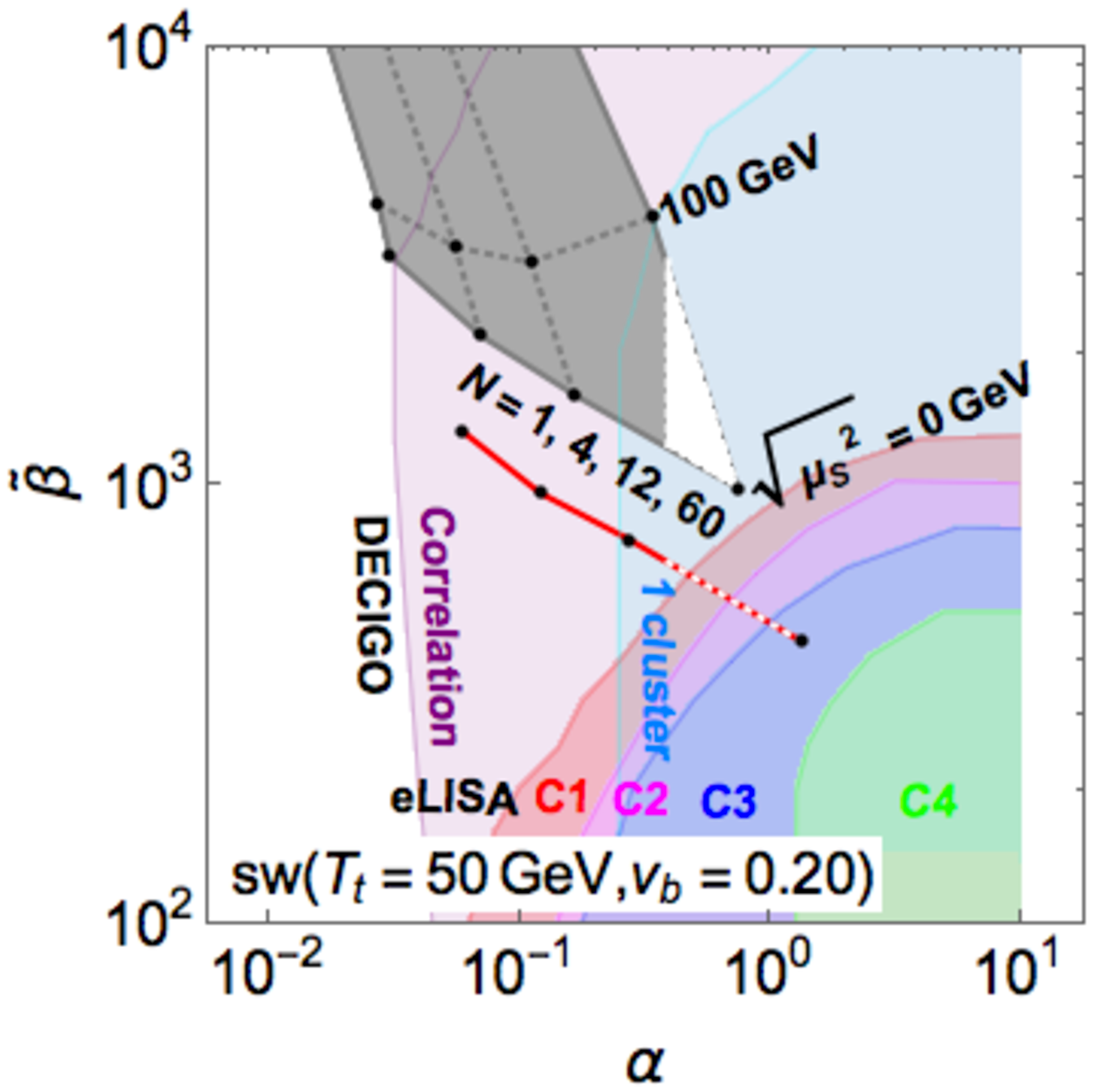}}
  \end{minipage}
  \begin{minipage}[t]{0.48\hsize}
  \centerline{\includegraphics[width=1\textwidth]{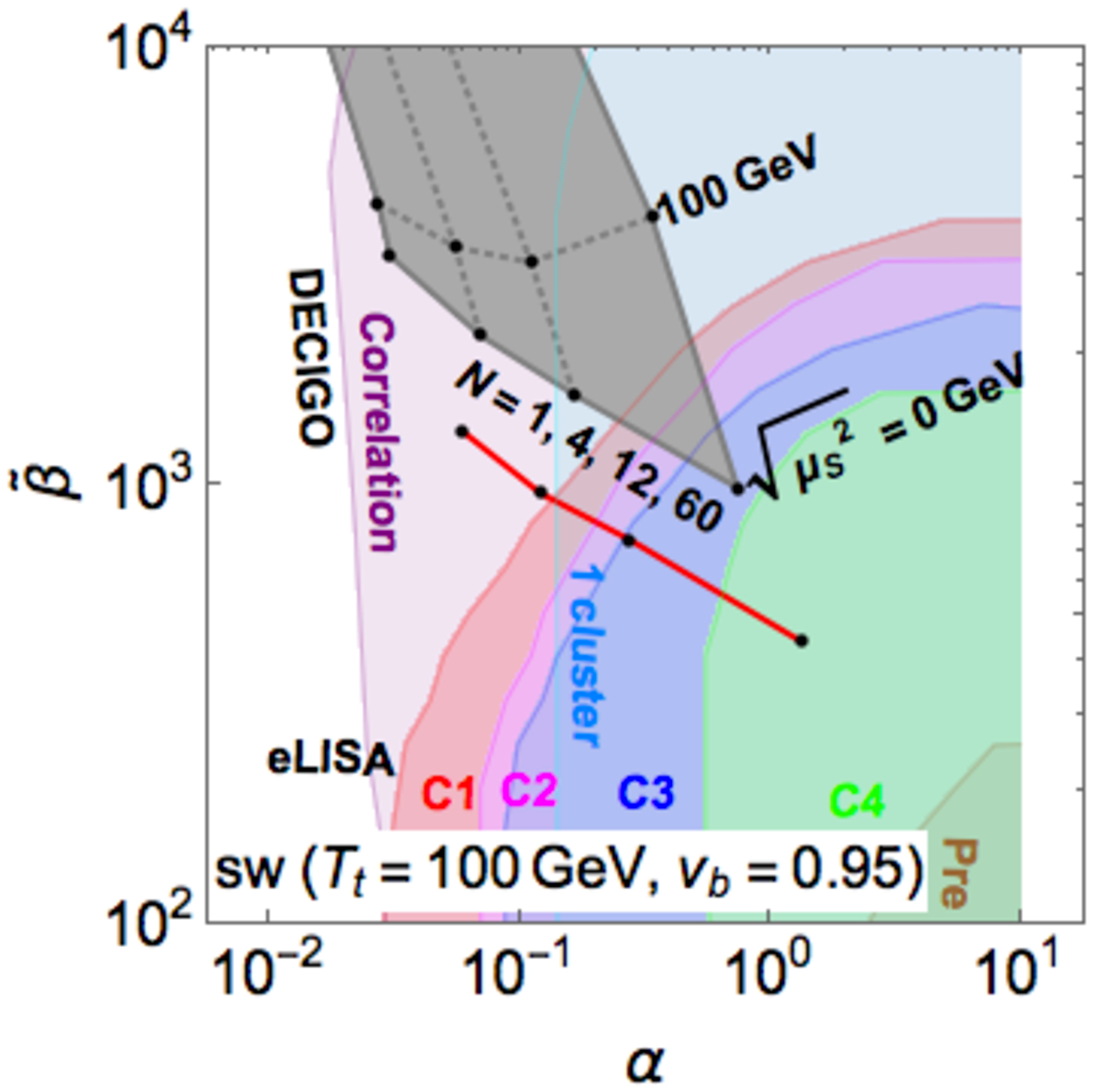}}
  \end{minipage}
  \begin{minipage}[t]{0.48\hsize}
  \centerline{\includegraphics[width=1\textwidth]{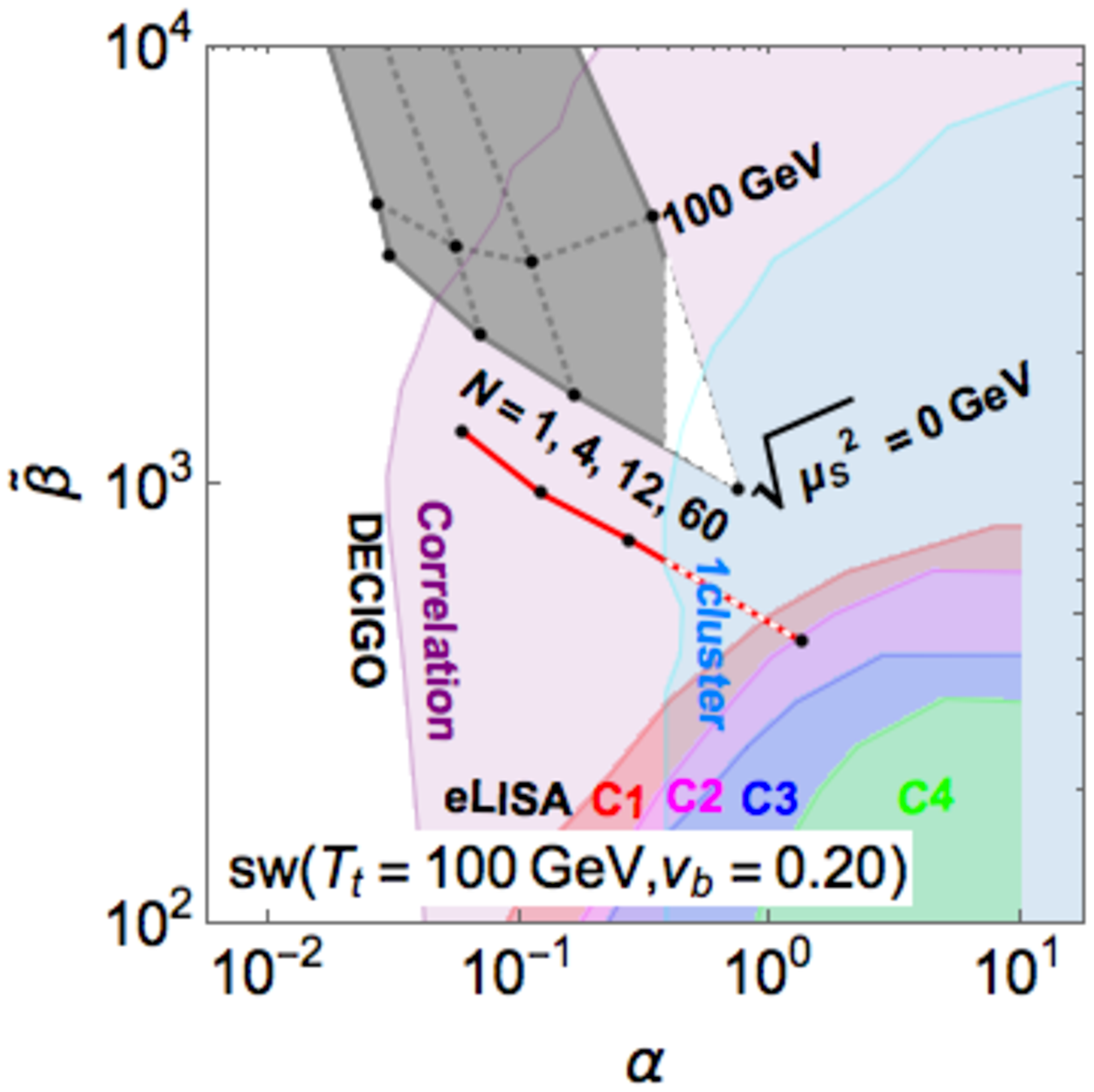}}
  \end{minipage}  
  \caption{
    Detectability of GWs in the CSI $O(N)$ models (red) and 
    $O(N)$ models without CSI with 
    $\Delta\lambda_{hhh}^{}/\lambda_{hhh}^{\SM}=2/3 (\simeq 70\%)$ (gray). 
    The experimental sensitivities expected 
    at the several designs of eLISA and DECIGO are
    set by using the sound wave contribution
    for $T_t^{}=50~\GeV$ (top frames) and $T_t^{}=100~\GeV$ (bottom frames).
    We take $v_b^{}=0.95$ (left) and $v_b^{}=0.2$ (right). 
    The upper bound on $\alpha$ ($\alpha=0.39$)
    is delineated for $v_b=0.2$ (right).
  }
  \label{fig:alphabeta_massless}
\end{figure}

In the case where the deviation in the $hhh$ coupling is found to be around
$70\%$ at electron-positron collider,
it is challenging for collider experiments to distinguish the CSI models from other 
extended models.
In Fig.~\ref{fig:alphabeta_massless}, the detectability of GWs 
in the CSI $O(N)$ models (red) 
and $O(N)$ model without CSI 
with $\Delta\lambda_{hhh}^{}/\lambda_{hhh}^{\SM}=2/3 (\simeq 70\%)$ (gray)
are displayed.
The experimental sensitivities expected 
at the several designs of eLISA and DECIGO are
set by using the sound wave contribution.
Actually, the transition temperature $T_t^{}$ differs according to models
as presented in Table~\ref{tab:ON}. 
For the purpose of illustration, 
the transition temperature is
fixed at $T_t^{}=50~\GeV$ (top) and $T_t^{}=100~\GeV$ (bottom).
As in Fig.~\ref{fig:GWspectrum}, we 
take $v_b^{}=0.95$ (left) and $v_b^{}=0.2$ (right).
Since the shape of the effective potentials of the CSI models 
is different from
those of the extended models without CSI,
their predicted regions in the ($\alpha$, $\beta$) plane do not overlap.
If the wall velocity is as large as $v_b^{}=0.95$,
eLISA C1 configuration can detect the produced GWs 
in the CSI models with $N \gtrsim 4$.
The sensitivity reach of the correlation program of DECIGO 
can cover all the CSI models even when the wall velocity is as small as 
$v_b^{}=0.2$.
Therefore, the two classes of model can be differentiated
at future GW detectors even if they share the common value for the $hhh$ coupling.

Before closing this section, we mention the predictions in the CSI two Higgs doublet model (2HDM)~\cite{Funakubo:1993jg,Takenaga:1993ux,Lee:2012jn,Fuyuto:2015vna}.
The Higgs sector of the 2HDM consists of two Higgs doublets
$\Phi_1$ and $\Phi_2$.
Therefore, the CSI 2HDM is
an explicit example corresponding to the CSI $O(4)$ model as the number of the extra scalars is common.
We have numerically 
checked that the predicted values of $\alpha$ and $\beta$ are
approximately equal to those in the CSI $O(4)$ model.

\clearpage
\section{Conclusion}
\label{sec:conclusion}

Nature of EWPT is important for understanding 
physics at the early Universe, such that the strongly 1stOPT is 
required for a successful scenario of EWBG. 
We have discussed complementarity of the methods of measurements of 
the $hhh$ coupling at the ILC and 
the spectrum of stochastic GWs at eLISA and DECIGO 
in testing the strongly 1stOPT of the EWSB in the $O(N)$ models 
with and without CSI. 
The models with CSI universally predict a large deviation in the 
$hhh$ coupling as large as about $70\%$, 
and leads to the strongly first order EWPT.
There is a possibility that
the deviation in the $hhh$ coupling
in the $O(N)$ models without CSI accidentally matches $70\%$.
Even in such a case, the predicted GW spectra are different with each other, 
and we can separate both the models by measuring GWs at future space-based interferometers. 
In conclusion, synergy between the future measurements of the $hhh$ coupling
and the GW signals provides us important hints regarding narrowing down
the dynamics behind the EWSB.

\begin{acknowledgments}
This work was supported, in part, by 
Grant-in-Aid for Scientific Research, No.\ 26104702 (MK) 
and No.\ 23104006 (SK), 
Grant H2020-MSCA-RISE-2014~no.~645722 (Non Minimal Higgs) (SK), 
and the Sasakawa Scientific Research Grant from The Japan Science Society (TM).
\end{acknowledgments}

\clearpage
\appendix
\section{Landau pole of the CSI $O(N)$ models}
\label{sec:landaupole}

We comment on the occurrence of the Landau pole in the CSI models.
Although values of the couplings in the scalar sector are 
in the perturbative range at around the electroweak scale,
large scalar couplings lead to the Landau pole 
at energy scales near the electroweak scale.
Therefore, some theory of UV completion should replace
the CSI models above this cutoff scale $\Lambda$. 

Let us estimate the energy scale $\Lambda$ of the Landau pole 
in the CSI $O(N)$ models by solving the renormalization group equations.
We derived the beta functions at one-loop level which are consistent with the ones in Ref.~\cite{Endo:2015ifa}. 
As for the input values of the couplings in the scalar sector,
we take $\lambda_{\Phi S}^{} = 2 m_S^2/v^2$ and $\lambda = 0$
at the boundary scale $Q$ fixed by Eq.~(\ref{eq:Q}).
The value of the singlet self-coupling $\lambda_S^{}$ is 
undetermined and constrained only by the unitarity bound.
In Table~\ref{tab:landau}, the energy scale of the the Landau pole $\Lambda$
is shown for $N=1, 4, 12$ and $60$.
Following the convention adopted in Ref.~\cite{Endo:2015ifa}, 
the cutoff scale $\Lambda$ in Table~\ref{tab:landau} is defined as
the scale where any of the scalar couplings diverges
\footnote{
  It may be a judicious choice to regard the Landau pole 
  as the scale where any of the scalar couplings reaches $\sim \!\!4 \pi$
  to keep perturbative expansion.
  In view of uncertainties involved 
  in the analysis of the renormalization group 
  flow, we disregard this discrepancy.
}. 
For $\lambda_S^{}>0.3$, the CSI model with $N=60$ satisfying the observed Higgs boson mass $m_h^{}=125~\GeV$ is excluded by the unitarity bound. 
As shown in the table, typically, 
the Landau pole appears at around ${\cal O}(10)$ TeV.
As $\lambda_S^{}$ is increased, the location of the Landau pole is
shifted to lower energy scales.
For moderate values of $N(\sim 10)$,
destructive interference 
between the top Yukawa and $\lambda_S$ contributions
is large, leading to higher cutoff scales.

\begin{table}
  \begin{center}
    \begin{tabular}{|l||c|c|c|c|}
      \hline
    $N$                & $1$       & $4$        & $12$     & $60$ \\ 
      \hline \hline
    $Q$    & 381\,GeV & 257\,GeV & 188\,GeV & 119\,GeV \\ 
      \hline
    $\Lambda(\lambda_S^{}=0)$    & 5.4\,TeV & 17\,TeV & 28\,TeV & 33\,TeV \\ 
      \hline
    $\Lambda(\lambda_S^{}=0.1)$  & 5.3\,TeV & 16\,TeV & 23\,TeV & 13\,TeV \\ 
      \hline
    $\Lambda(\lambda_S^{}=0.2)$  & 5.2\,TeV & 15\,TeV & 19\,TeV & 5.4\,TeV \\ 
      \hline
    $\Lambda(\lambda_S^{}=0.3)$  & 5.0\,TeV & 14\,TeV & 15\,TeV & 2.7\,TeV \\ 
      \hline
  \end{tabular}
  \end{center}
  \caption{The energy scale of the Landau pole $\Lambda$
    in the CSI $O(N)$ models for $N=1, 4, 12$ and $60$.}
  \label{tab:landau}
\end{table}

\section{Comparison between the recent and conventional analyses of the gravitational spectrum}
\label{sec:comparison}

\begin{figure}[t]
%
\begin{minipage}[t]{0.48\hsize}
  \centerline{\includegraphics[width=1\textwidth]{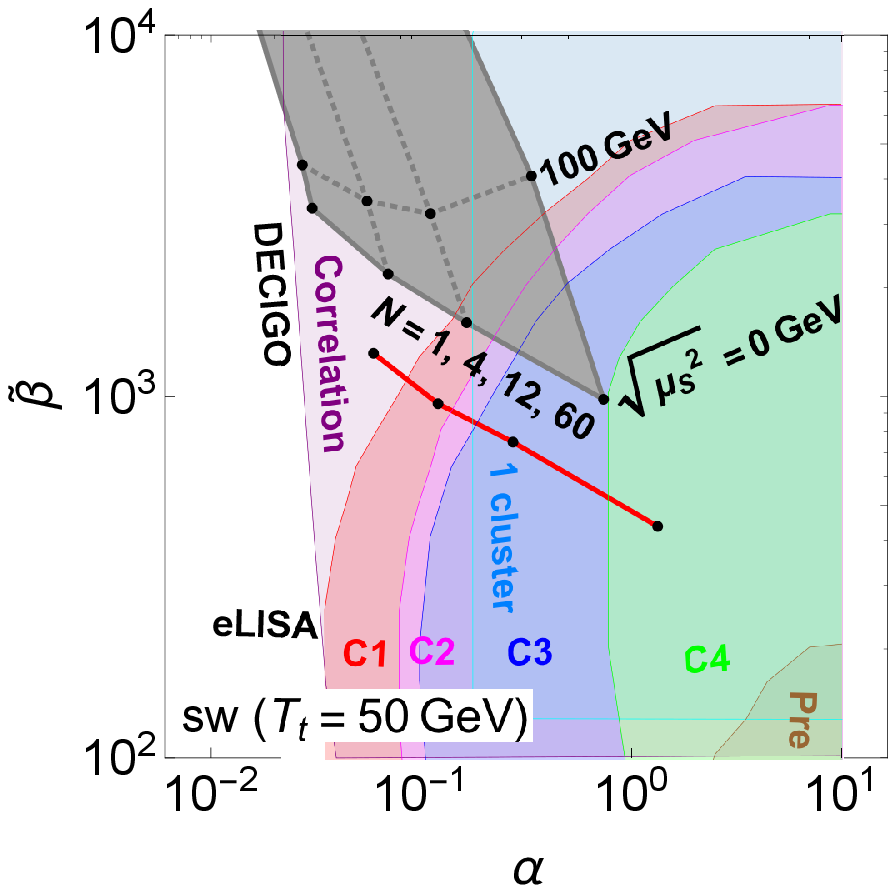}}
 \end{minipage}
\hfill
\begin{minipage}[t]{0.48\hsize}
 \centerline{\includegraphics[width=1\textwidth]{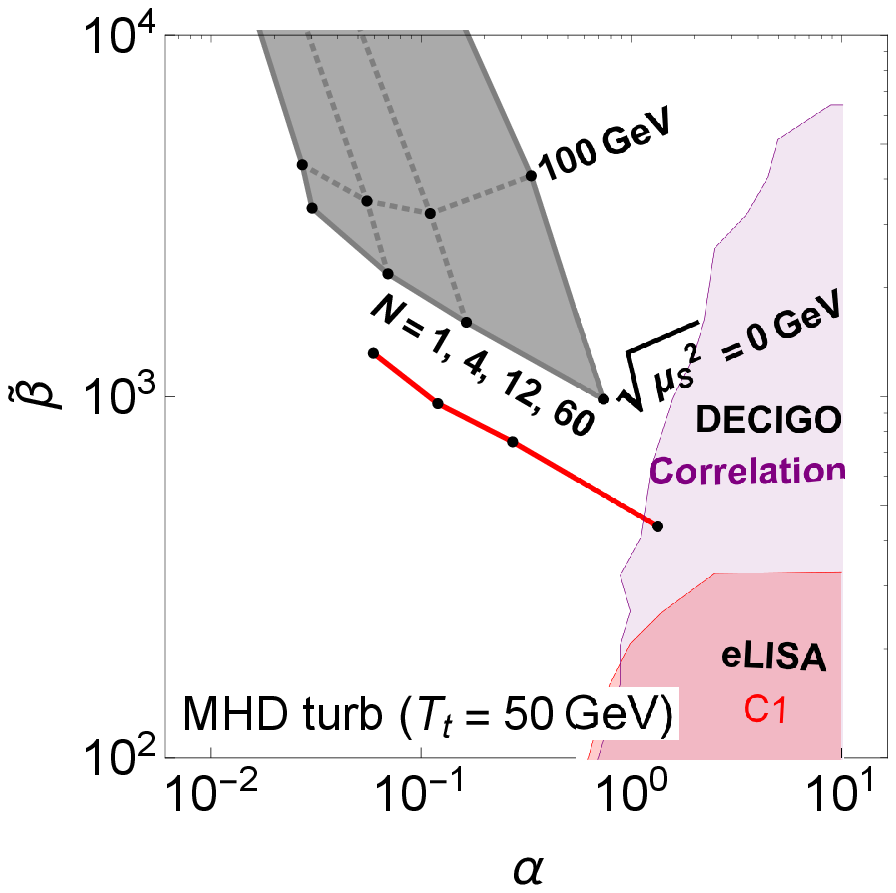}}
    \end{minipage}
\caption{
The experimental sensitivity in the $(\alpha, \tilde{\beta})$ plane based on the recent analysis given in Ref.~\cite{Caprini:2015zlo}. We show the contributions from 
sound waves (left) and MHD turbulence (right) assuming $T_t^{}=50~\GeV$ and $v_b=0.95$. 
}
\label{fig:alphabeta_new}
\end{figure}

In Fig.~\ref{fig:alphabeta_new}, we present the experimental sensitivity to the contributions from 
MHD turbulence in addition to the sound waves at $T_t^{}=50~\GeV$. 
As illustrated, we see that the sound waves are the dominant contribution. 

\begin{figure}[t]
\begin{minipage}[t]{0.48\hsize}
  \centerline{\includegraphics[width=1\textwidth]{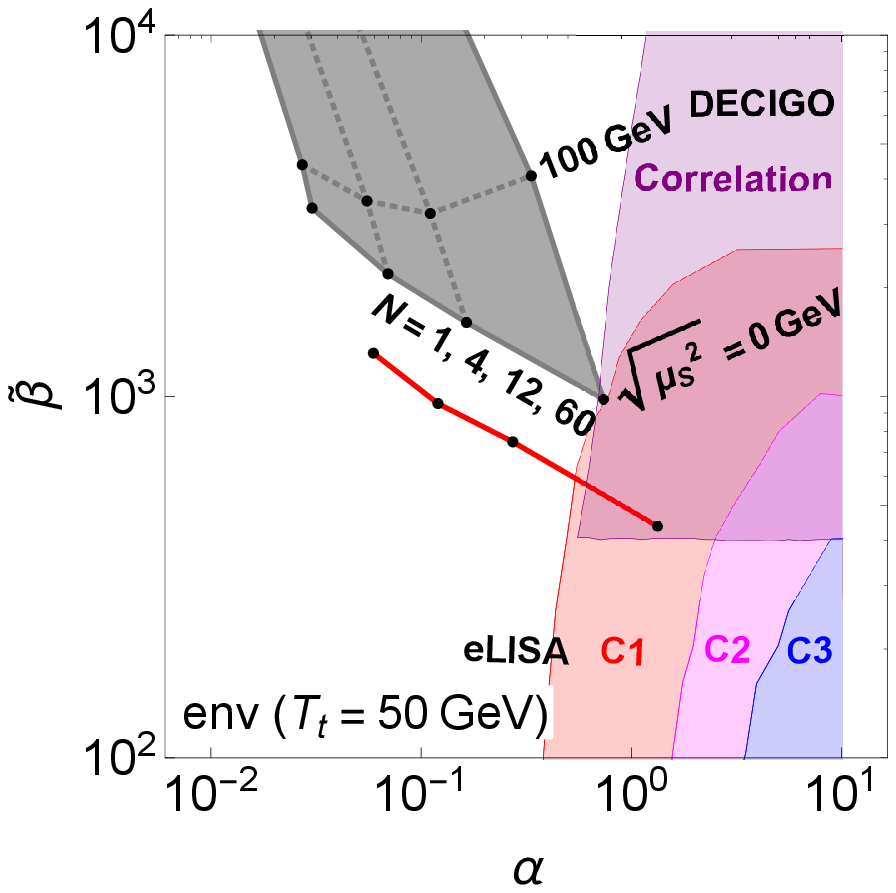}}
    \end{minipage}
\hfill
\begin{minipage}[t]{0.48\hsize}
  \centerline{\includegraphics[width=1\textwidth]{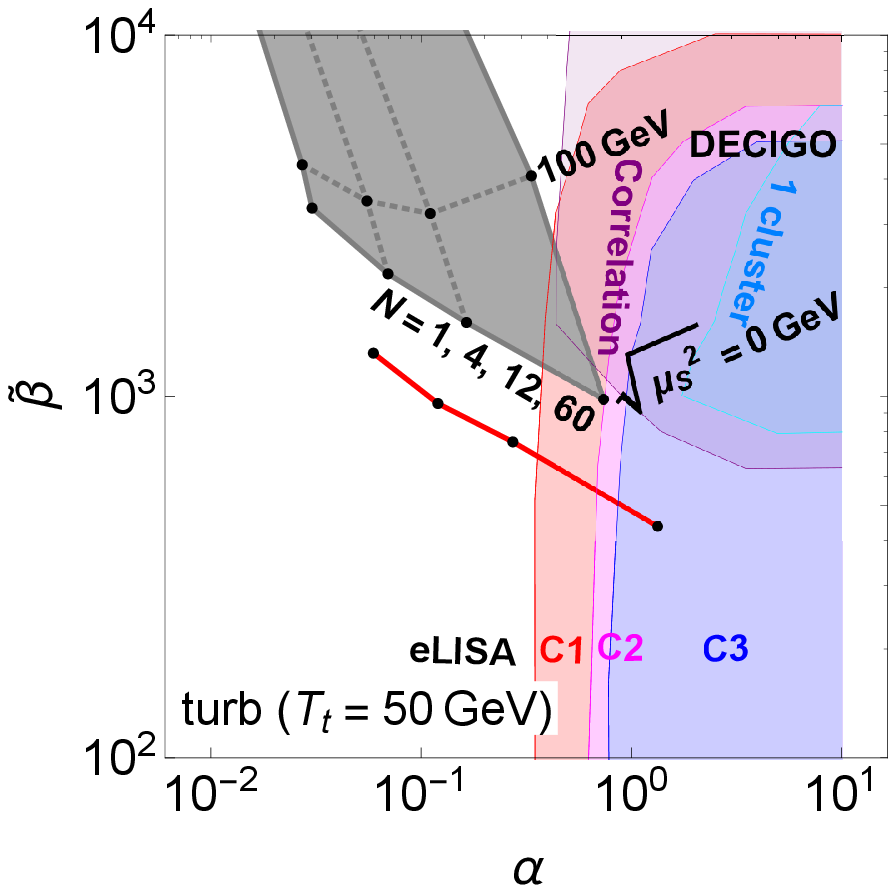}}
    \end{minipage}
  \caption{
    The experimental sensitivity in the $(\alpha, \tilde{\beta})$ plane based on the conventional analysis given in Ref.~\cite{Grojean:2006bp}. 
    We show the contributions from the collision of bubble walls using the envelope approximation (left) and turbulence (right) assuming $T_t^{}=50~\GeV$. 
  }
  \label{fig:alphabeta_old}
\end{figure}

Next, we compare these results based on the recent work given in Ref.~\cite{Caprini:2015zlo} with the estimate of the GW spectrum based on Ref.~\cite{Grojean:2006bp} which we employed in our previous work~\cite{Kakizaki:2015wua}. 
Results based on the conventional analysis given in Ref.~\cite{Grojean:2006bp} is shown in Fig.~\ref{fig:alphabeta_old}. 
As for the contribution from the turbulence in the plasma, 
the difference between Fig.~\ref{fig:alphabeta_new} (right) and Fig.~\ref{fig:alphabeta_old} (right) is attributed to MHD effects~\cite{Huber:2007vva}. 

\begin{figure}[t]
  \begin{minipage}[t]{0.48\hsize}
    \centerline{\includegraphics[width=1\textwidth]{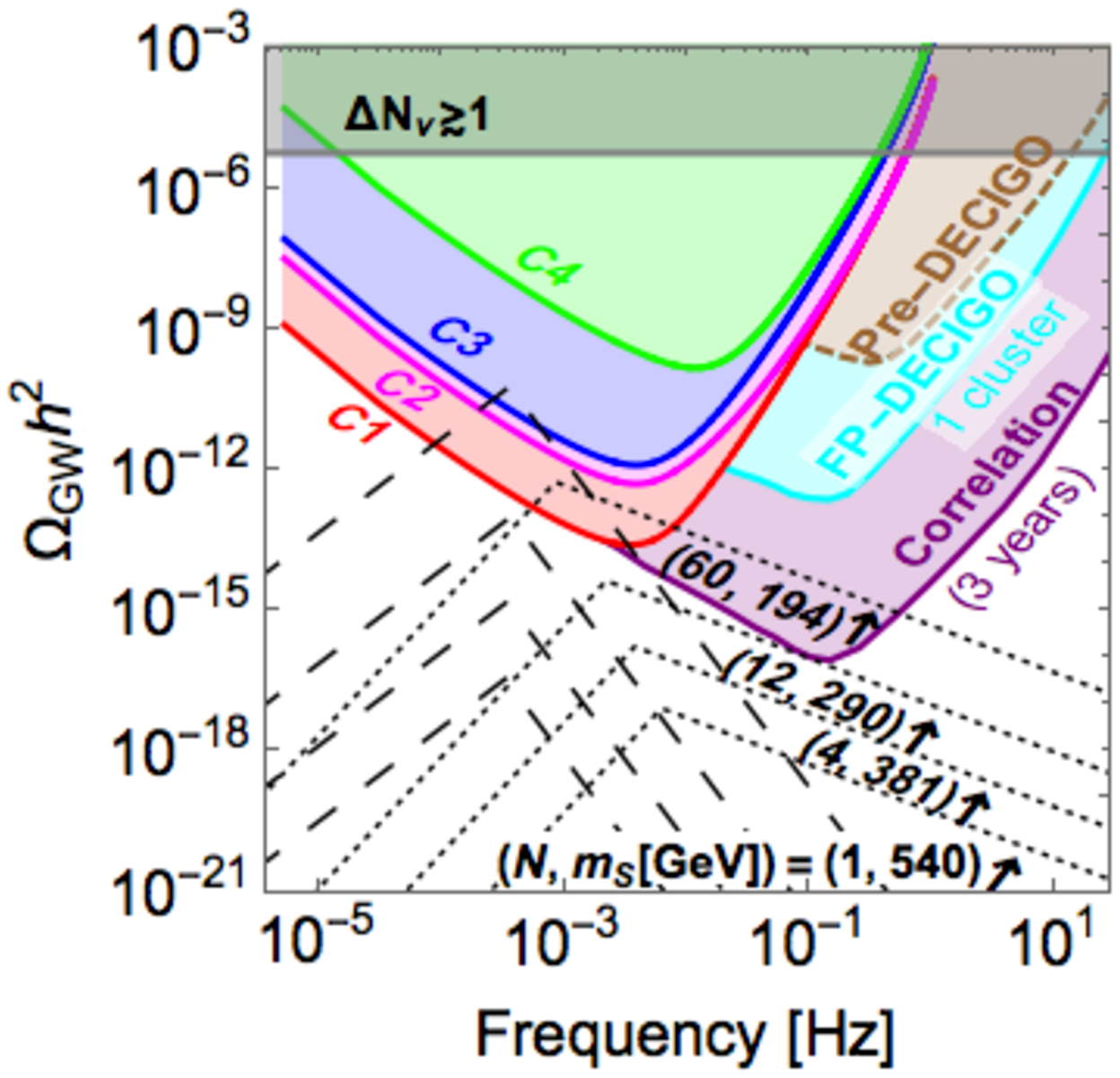}}
  \end{minipage}
  \begin{minipage}[t]{0.48\hsize}
    \centerline{\includegraphics[width=1\textwidth]{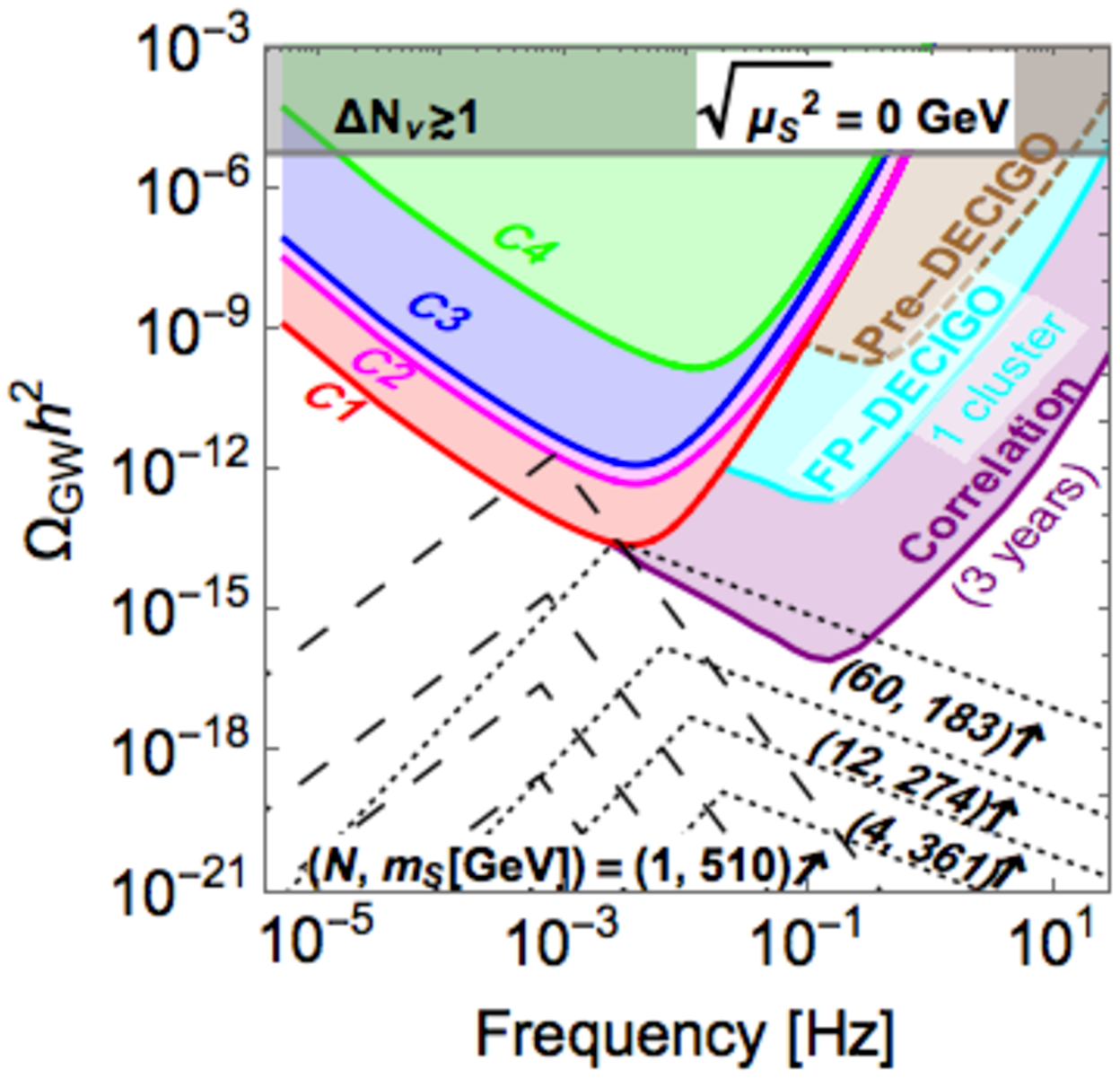}}
  \end{minipage}
  \caption{
    GW spectra in the CSI $O(N)$ models (left) 
    and the $O(N)$ models without CSI 
    with $\Delta\lambda_{hhh}^{}/\lambda_{hhh}^{\SM}=2/3 (\simeq 70\%)$ (right) 
    based on the conventional analysis given in Ref.~\cite{Grojean:2006bp}. 
    The black curves correspond 
    to the contributions from the bubble collision (dotted)
    and turbulence (dashed) 
    for $N=1,4,12$ and $60$ from the bottom.
  }
  \label{fig:GWspectrum_old}
\end{figure}

In Fig.~\ref{fig:GWspectrum_old}, we show GW spectra based on the conventional analysis given in Ref.~\cite{Grojean:2006bp}, 
since the sound wave contribution is absent, the predicted GWs are relatively weaker compared to those shown in Fig.~\ref{fig:GWspectrum}. 



\begin{thebibliography}{99}

\bibitem{Aad:2012tfa} 
  G.~Aad {\it et al.}  [ATLAS Collaboration],
  Phys.\ Lett.\ B {\bf 716}, 1 (2012). 

\bibitem{Chatrchyan:2012xdj} 
  S.~Chatrchyan {\it et al.} [CMS Collaboration],
  Phys.\ Lett.\ B {\bf 716}, 30 (2012).

\bibitem{Khachatryan:2014jba} 
  V.~Khachatryan {\it et al.} [CMS Collaboration],
  Eur.\ Phys.\ J.\ C {\bf 75}, no. 5, 212 (2015). 

\bibitem{Aad:2015gba} 
  G.~Aad {\it et al.} [ATLAS Collaboration],
  Eur.\ Phys.\ J.\ C {\bf 76}, no. 1, 6 (2016). 

\bibitem{Agashe:2014kda}
  K.~A.~Olive {\it et al.} [Particle Data Group Collaboration],
  Chin.\ Phys.\ C {\bf 38}, 090001 (2014).

\bibitem{Kuzmin:1985mm} 
  V.~A.~Kuzmin, V.~A.~Rubakov and M.~E.~Shaposhnikov,
  Phys.\ Lett.\ B {\bf 155}, 36 (1985).

\bibitem{Funakubo:1993jg} 
  K.~Funakubo, A.~Kakuto and K.~Takenaga,
  Prog.\ Theor.\ Phys.\  {\bf 91}, 341 (1994). 

\bibitem{Cline:1996mga} 
  J.~M.~Cline and P.~A.~Lemieux,
  Phys.\ Rev.\ D {\bf 55}, 3873 (1997). 

\bibitem{Kanemura:2004ch} 
  S.~Kanemura, Y.~Okada and E.~Senaha,
  Phys.\ Lett.\ B {\bf 606}, 361 (2005).

\bibitem{Funakubo:2005pu} 
  K.~Funakubo, S.~Tao and F.~Toyoda,
  Prog.\ Theor.\ Phys.\  {\bf 114}, 369 (2005). 

\bibitem{Profumo:2007wc} 
  S.~Profumo, M.~J.~Ramsey-Musolf and G.~Shaughnessy,
  JHEP {\bf 0708}, 010 (2007). 

\bibitem{Noble:2007kk} 
  A.~Noble and M.~Perelstein,
  Phys.\ Rev.\ D {\bf 78}, 063518 (2008). 

\bibitem{Aoki:2008av} 
  M.~Aoki, S.~Kanemura and O.~Seto,
  Phys.\ Rev.\ Lett.\  {\bf 102}, 051805 (2009). 

\bibitem{Kanemura:2011fy} 
  S.~Kanemura, E.~Senaha and T.~Shindou,
  Phys.\ Lett.\ B {\bf 706}, 40 (2011). 

\bibitem{Gil:2012ya} 
  G.~Gil, P.~Chankowski and M.~Krawczyk,
  Phys.\ Lett.\ B {\bf 717}, 396 (2012). 

\bibitem{Tamarit:2014dua} 
  C.~Tamarit,
  Phys.\ Rev.\ D {\bf 90}, no. 5, 055024 (2014). 

\bibitem{Kanemura:2014cka} 
  S.~Kanemura, N.~Machida and T.~Shindou,
  Phys.\ Lett.\ B {\bf 738}, 178 (2014). 

\bibitem{Fuyuto:2014yia} 
  K.~Fuyuto and E.~Senaha,
  Phys.\ Rev.\ D {\bf 90}, no. 1, 015015 (2014). 

\bibitem{Profumo:2014opa} 
  S.~Profumo, M.~J.~Ramsey-Musolf, C.~L.~Wainwright and P.~Winslow,
  Phys.\ Rev.\ D {\bf 91}, no. 3, 035018 (2015). 

\bibitem{Blinov:2015vma} 
  N.~Blinov, S.~Profumo and T.~Stefaniak,
  JCAP {\bf 1507}, no. 07, 028 (2015). 

\bibitem{Fuyuto:2015vna} 
  K.~Fuyuto and E.~Senaha,
  Phys.\ Lett.\ B {\bf 747}, 152 (2015).

\bibitem{Karam:2015jta} 
  A.~Karam and K.~Tamvakis,
  Phys.\ Rev.\ D {\bf 92}, no. 7, 075010 (2015). 

\bibitem{Kakizaki:2015wua} 
  M.~Kakizaki, S.~Kanemura and T.~Matsui,
  Phys.\ Rev.\ D {\bf 92}, no. 11, 115007 (2015). 

\bibitem{Grojean:2004xa} 
  C.~Grojean, G.~Servant and J.~D.~Wells,
  Phys.\ Rev.\ D {\bf 71}, 036001 (2005). 

\bibitem{Grinstein:2008qi} 
  B.~Grinstein and M.~Trott,
  Phys.\ Rev.\ D {\bf 78}, 075022 (2008). 

\bibitem{ILC}
%
  J.~Brau, (Ed.) {\it et al.}  [ILC Collaboration],
  arXiv:0712.1950 [physics.acc-ph];
%
  G.~Aarons {\it et al.}  [ILC Collaboration],
  arXiv:0709.1893 [hep-ph];
%
  N.~Phinney, N.~Toge and N.~Walker,
  arXiv:0712.2361 [physics.acc-ph];
%
  T.~Behnke, (Ed.) {\it et al.}  [ILC Collaboration],
  arXiv:0712.2356 [physics.ins-det];
%
T.~Behnke {\it et al.},
arXiv:1306.6329 [physics.ins-det];
%
H.~Baer, {\it et al.} "Physics at the International Linear Collider", 
{\it Physics Chapter of the ILC Detailed Baseline Design Report}:
http://lcsim.org/papers/DBDPhysics.pdf.

\bibitem{ILCHiggsWhitePaper} 
  D.~M.~Asner, T.~Barklow, C.~Calancha, K.~Fujii, N.~Graf, H.~E.~Haber, A.~Ishikawa and S.~Kanemura {\it et al.},
  arXiv:1310.0763 [hep-ph].

\bibitem{Fujii:2015jha} 
  K.~Fujii {\it et al.},
  arXiv:1506.05992 [hep-ex].

\bibitem{CLIC}
  E.~Accomando {\it et al.}  [CLIC Physics Working Group Collaboration],
  hep-ph/0412251;
  L.~Linssen, A.~Miyamoto, M.~Stanitzki and H.~Weerts,
  arXiv:1202.5940 [physics.ins-det].

\bibitem{FCC-ee}
  M.~Bicer {\it et al.} [TLEP Design Study Working Group Collaboration],
  JHEP {\bf 1401}, 164 (2014). 

\bibitem{He:2015spf} 
  H.~J.~He, J.~Ren and W.~Yao,
  Phys.\ Rev.\ D {\bf 93}, no. 1, 015003 (2016). 

\bibitem{Grojean:2006bp} 
  C.~Grojean and G.~Servant,
  Phys.\ Rev.\ D {\bf 75}, 043507 (2007).

\bibitem{Abbott:2016blz}
  B.~P.~Abbott {\it et al.} [LIGO Scientific and Virgo Collaborations],
  Phys.\ Rev.\ Lett.\  {\bf 116}, no. 6, 061102 (2016). 

\bibitem{Somiya:2011np} 
  K.~Somiya  [KAGRA Collaboration],
  Class.\ Quant.\ Grav.\  {\bf 29}, 124007 (2012).

\bibitem{Harry:2010zz} 
  G.~M.~Harry [LIGO Scientific Collaboration],
  Class.\ Quant.\ Grav.\  {\bf 27}, 084006 (2010).

\bibitem{Accadia:2009zz} 
  T.~Accadia {\it et al.}, 
  Proceedings of {\it 12th Marcel Grossmann Meeting on General Relativity}, 
  pp.~1738--1742 (2009).

\bibitem{Seoane:2013qna} 
  P.~A.~Seoane {\it et al.} [eLISA Collaboration],
  arXiv:1305.5720 [astro-ph.CO].

\bibitem{Kawamura:2011zz} 
  S.~Kawamura {\it et al.},
  Class.\ Quant.\ Grav.\  {\bf 28}, 094011 (2011).
  
\bibitem{Corbin:2005ny} 
  V.~Corbin and N.~J.~Cornish,
  Class.\ Quant.\ Grav.\  {\bf 23}, 2435 (2006).

\bibitem{Kamionkowski:1993fg} 
  M.~Kamionkowski, A.~Kosowsky and M.~S.~Turner,
  Phys.\ Rev.\ D {\bf 49}, 2837 (1994).

\bibitem{Huber:2008hg} 
  S.~J.~Huber and T.~Konstandin,
  JCAP {\bf 0809}, 022 (2008). 
  
\bibitem{Espinosa:2010hh} 
  J.~R.~Espinosa, T.~Konstandin, J.~M.~No and G.~Servant,
  JCAP {\bf 1006}, 028 (2010).

\bibitem{No:2011fi} 
  J.~M.~No,
  Phys.\ Rev.\ D {\bf 84}, 124025 (2011).
  
\bibitem{Hindmarsh:2013xza} 
  M.~Hindmarsh, S.~J.~Huber, K.~Rummukainen and D.~J.~Weir,
  Phys.\ Rev.\ Lett.\  {\bf 112}, 041301 (2014). 

\bibitem{Hindmarsh:2015qta} 
  M.~Hindmarsh, S.~J.~Huber, K.~Rummukainen and D.~J.~Weir,
  arXiv:1504.03291 [astro-ph.CO].

\bibitem{Caprini:2015zlo} 
  C.~Caprini {\it et al.},
  arXiv:1512.06239 [astro-ph.CO].

\bibitem{Leitao:2012tx} 
  L.~Leitao, A.~Megevand and A.~D.~Sanchez,
  JCAP {\bf 1210}, 024 (2012). 

\bibitem{Kikuta:2014eja} 
  Y.~Kikuta, K.~Kohri and E.~So,
  arXiv:1405.4166 [hep-ph].

\bibitem{Jinno:2015doa} 
  R.~Jinno, K.~Nakayama and M.~Takimoto,
  Phys.\ Rev.\ D {\bf 93}, no. 4, 045024 (2016). 

\bibitem{Leitao:2015fmj} 
  L.~Leitao and A.~Megevand,
  arXiv:1512.08962 [astro-ph.CO].

\bibitem{Jaeckel:2016jlh} 
  J.~Jaeckel, V.~V.~Khoze and M.~Spannowsky,
  arXiv:1602.03901 [hep-ph].

\bibitem{Dev:2016feu} 
  P.~S.~B.~Dev and A.~Mazumdar,
  arXiv:1602.04203 [hep-ph].

\bibitem{Delaunay:2007wb} 
  C.~Delaunay, C.~Grojean and J.~D.~Wells,
  JHEP {\bf 0804}, 029 (2008).

\bibitem{Huang:2016odd} 
  F.~P.~Huang, Y.~Wan, D.~G.~Wang, Y.~F.~Cai and X.~Zhang,
  arXiv:1601.01640 [hep-ph].

\bibitem{Espinosa:2007qk} 
  J.~R.~Espinosa and M.~Quiros,
  Phys.\ Rev.\ D {\bf 76}, 076004 (2007).

\bibitem{Espinosa:2008kw} 
  J.~R.~Espinosa, T.~Konstandin, J.~M.~No and M.~Quiros,
  Phys.\ Rev.\ D {\bf 78}, 123528 (2008).
  
\bibitem{Apreda:2001us} 
  R.~Apreda, M.~Maggiore, A.~Nicolis and A.~Riotto,
  Nucl.\ Phys.\ B {\bf 631}, 342 (2002).

\bibitem{Ashoorioon:2009nf} 
  A.~Ashoorioon and T.~Konstandin,
  JHEP {\bf 0907}, 086 (2009).

\bibitem{Sagunski:2012ufa} 
  L.~Sagunski,
  DESY-THESIS-2013-011.

\bibitem{Huber:2015znp} 
  S.~J.~Huber, T.~Konstandin, G.~Nardini and I.~Rues,
  arXiv:1512.06357 [hep-ph].

\bibitem{Bardeen}
W. A. Bardeen, FERMILAB-CONF-95-391-T.

\bibitem{Coleman:1973jx} 
  S.~R.~Coleman and E.~J.~Weinberg,
  Phys.\ Rev.\ D {\bf 7}, 1888 (1973).

\bibitem{Gildener:1976ih} 
  E.~Gildener and S.~Weinberg,
  Phys.\ Rev.\ D {\bf 13}, 3333 (1976).

\bibitem{Takenaga:1993ux} 
  K.~Takenaga,
  Prog.\ Theor.\ Phys.\  {\bf 92}, 987 (1994); 
  {\bf 95}, 609 (1996). 

\bibitem{Lee:2012jn} 
  J.~S.~Lee and A.~Pilaftsis,
  Phys.\ Rev.\ D {\bf 86}, 035004 (2012).

\bibitem{Endo:2015ifa} 
  K.~Endo and Y.~Sumino,
  JHEP {\bf 1505}, 030 (2015). 

\bibitem{Endo:2015nba} 
  K.~Endo and K.~Ishiwata,
  Phys.\ Lett.\ B {\bf 749}, 583 (2015). 

\bibitem{Plascencia:2015xwa} 
  A.~D.~Plascencia,
  JHEP {\bf 1509}, 026 (2015). 

\bibitem{Hashino:2015nxa} 
  K.~Hashino, S.~Kanemura and Y.~Orikasa,
  Phys.\ Lett.\ B {\bf 752}, 217 (2016). 

\bibitem{Endo:2016koi} 
  K.~Endo, K.~Ishiwata and Y.~Sumino,
  arXiv:1601.00696 [hep-ph].

\bibitem{DRbar}
  W.~Siegel,
  Phys.\ Lett.\ B {\bf 84}, 193 (1979); 
{\bf 94}, 37 (1980); 
  D.~M.~Capper, D.~R.~T.~Jones and P.~van Nieuwenhuizen,
  Nucl.\ Phys.\ B {\bf 167}, 479 (1980).

\bibitem{Kanemura:2002vm} 
  S.~Kanemura, S.~Kiyoura, Y.~Okada, E.~Senaha and C.~P.~Yuan,
  Phys.\ Lett.\ B {\bf 558}, 157 (2003). 

\bibitem{Kanemura:2004mg} 
  S.~Kanemura, Y.~Okada, E.~Senaha and C.-P.~Yuan,
  Phys.\ Rev.\ D {\bf 70}, 115002 (2004). 

\bibitem{LHChhh}
  CMS Collaboration [CMS Collaboration],
  CMS-PAS-FTR-15-002.

\bibitem{Dolan:1973qd} 
  L.~Dolan and R.~Jackiw,
  Phys.\ Rev.\ D {\bf 9}, 3320 (1974).

\bibitem{Carrington:1991hz} 
  M.~E.~Carrington,
  Phys.\ Rev.\ D {\bf 45}, 2933 (1992).

\bibitem{Sakharov:1967dj} 
  A.~D.~Sakharov,
  Pisma Zh.\ Eksp.\ Teor.\ Fiz.\  {\bf 5}, 32 (1967).

\bibitem{Caprini:2009yp} 
  C.~Caprini, R.~Durrer and G.~Servant,
  JCAP {\bf 0912}, 024 (2009). 

\bibitem{Nicolis:2003tg} 
  A.~Nicolis,
  Class.\ Quant.\ Grav.\  {\bf 21}, L27 (2004).

\bibitem{Binetruy:2012ze} 
  P.~Binetruy, A.~Bohe, C.~Caprini and J.~F.~Dufaux,
  JCAP {\bf 1206}, 027 (2012). 

\bibitem{Klein:2015hvg} 
  A.~Klein {\it et al.},
  Phys.\ Rev.\ D {\bf 93}, no. 2, 024003 (2016). 

\bibitem{PetiteauDataSheet}
Data sheet by A. Petiteau, \\
http://www.apc.univ-paris7.fr/Downloads/lisa/eLISA/Sensitivity/Cfgv1/StochBkgd/

\bibitem{Joyce:1994fu} 
  M.~Joyce, T.~Prokopec and N.~Turok,
  Phys.\ Rev.\ Lett.\  {\bf 75}, 1695 (1995); 
%
  Phys.\ Rev.\ D {\bf 53}, 2930 (1996); 
%
  {\bf 53}, 2958 (1996). 

\bibitem{planck}
  P.~A.~R.~Ade {\it et al.}  [Planck Collaboration],
  arXiv:1502.01589 [astro-ph.CO].

\bibitem{Schneider:2010ks} 
  R.~Schneider, S.~Marassi and V.~Ferrari,
  Class.\ Quant.\ Grav.\  {\bf 27}, 194007 (2010).

\bibitem{Nelemans:2009hy} 
  G.~Nelemans,
  Class.\ Quant.\ Grav.\  {\bf 26}, 094030 (2009). 

\bibitem{Huber:2007vva} 
  S.~J.~Huber and T.~Konstandin,
  JCAP {\bf 0805}, 017 (2008). 

\end{thebibliography}
\end{document}